\documentclass[9pt,sigconf,screen=true,bookmarks=false]{acmart}
 
\usepackage{color,xcolor}
\usepackage{epsfig}
\usepackage{graphicx}

\usepackage{adjustbox}
\usepackage{array}
\usepackage{booktabs}
\usepackage{colortbl}
\usepackage{float,wrapfig}
\usepackage{hhline}
\usepackage{multirow}
\usepackage{amsfonts}
\usepackage{mathtools}
\usepackage{bm}
\usepackage{nicefrac}
\usepackage{braket}

\usepackage{changepage}
\usepackage{extramarks}
\usepackage{fancyhdr}
\usepackage{setspace}
\usepackage{soul}
\usepackage{xspace}
\usepackage{textcomp}
\usepackage[subrefformat=parens,labelformat=parens]{subfig}

\usepackage{enumerate}
\usepackage{fancyhdr,graphicx,amsmath}
\usepackage[ruled,vlined]{algorithm2e}
\usepackage{optidef}
\usepackage{enumitem}

\definecolor{citecolor}{RGB}{34,139,34}
\definecolor{mydarkblue}{rgb}{0,0.08,1}
\definecolor{mydarkgreen}{rgb}{0.02,0.6,0.02}
\definecolor{mydarkred}{rgb}{0.8,0.02,0.02}
\definecolor{mydarkorange}{rgb}{0.40,0.2,0.02}
\definecolor{mypurple}{RGB}{111,0,255}
\definecolor{myred}{rgb}{1.0,0.0,0.0}
\definecolor{mygold}{rgb}{0.75,0.6,0.12}
\definecolor{myblue}{rgb}{0,0.2,0.8}
\definecolor{mydarkgray}{rgb}{0.,0.2,0.2}

\definecolor{lightred}{RGB}{255,235,235}
\definecolor{lightgreen}{RGB}{235,255,235}
\definecolor{lightblue}{RGB}{235,235,255}
\definecolor{lightcyan}{RGB}{235,255,255}
\definecolor{lightmagenta}{RGB}{255,235,255}
\definecolor{lightyellow}{RGB}{255,255,235}

\definecolor{qxkcolor}{RGB}{215,235,255}
\definecolor{softmaxcolor}{RGB}{230,235,255}
\definecolor{probxvcolor}{RGB}{255,255,235}

\definecolor{topkcolor}{RGB}{255,235,235}
\definecolor{zecolor}{RGB}{255,255,235}
\definecolor{dynacolor}{RGB}{235,255,255}

\definecolor{reviewcolor}{RGB}{0,0,200}

\renewcommand\footnotemark{}
\newcommand{\name}{Q-Pilot\xspace}

\newcommand{\fpqa}{FPQA\xspace}

\newcommand{\etal}{\emph{et al.}\xspace}

\newcommand{\cnot}{\texttt{CNOT}}

\newcommand{\swap}{\texttt{SWAP}}

\newcommand{\zz}{\texttt{ZZ}}
\newcommand{\cz}{\texttt{CZ}}
\newcommand{\cx}{\texttt{CNOT}}

\newcommand{\raa}{FPQA\xspace}

\newcommand{\fpqafull}{Field Programmable Qubit Array\xspace}

\newcounter{rlabelno}

\usepackage{geometry}

\expandafter\def\expandafter\normalsize\expandafter{%
    \normalsize%
    \setlength\abovedisplayskip{2pt}%
    \setlength\belowdisplayskip{2pt}%
    \setlength\abovedisplayshortskip{2pt}%
    \setlength\belowdisplayshortskip{2pt}%
}

\copyrightyear{2024} 
\acmYear{2024} 
\setcopyright{rightsretained} 
\acmConference[DAC '24]{61st ACM/IEEE Design Automation Conference}{June 23--27, 2024}{San Francisco, CA, USA}
\acmBooktitle{61st ACM/IEEE Design Automation Conference (DAC '24), June 23--27, 2024, San Francisco, CA, USA}
\acmDOI{10.1145/3649329.3658470}
\acmISBN{979-8-4007-0601-1/24/06}

\begin{document}
\settopmatter{printacmref=false} % Removes citation information below abstract
% \bstctlcite{IEEEexample:BSTcontrol}
% \renewcommand\footnotetextcopyrightpermission[1]{} % removes footnote with conference information in first column

%\settopmatter{printacmref=false} % Removes citation information below abstract
%\renewcommand\footnotetextcopyrightpermission[1]{} % removes footnote with conference information in first column
% \pagestyle{plain} % removes running headers

% \pagestyle{fancy}
% \fancyhead[L]{}
% \fancyhead[R]{}
% \fancypagestyle{firstpage}{%
%   \chead{The 59th Design Automation Conference (DAC 2022)}
% %   \rhead{**Right Header for just the first page**}
% }

% \title{
% RobustQNN: Noise-Aware Training for \\ Robust Quantum Neural Networks
% }

% \title{
% RobustQNN: Noise-Aware Training for \\ Robust Quantum Neural Networks
% }

\newcommand{\equal}[1]{{\hypersetup{linkcolor=black}\thanks{#1}}}

\pagestyle{plain}

\title{\huge \name: Field Programmable Qubit Array Compilation with Flying Ancillas}
\author{
Hanrui Wang$^{1*}$, Daniel Bochen Tan$^{2*}$, Pengyu Liu$^3$, Yilian Liu$^4$, Jiaqi Gu$^5$, Jason Cong$^2$, Song Han$^1$\\
\footnotesize $^1$MIT, $^2$University of California, Los Angeles, $^3$Carnegie Mellon University, $^4$Cornell University, $^5$Arizona State University, *Equal Contributions\\ 
}

\begin{abstract}

Neutral atom arrays have become a promising platform for quantum computing, especially the \textit{field programmable qubit array} (FPQA) endowed with the unique capability of atom movement.
This feature allows dynamic alterations in qubit connectivity during runtime, which can reduce the cost of executing long-range gates and improve parallelism.
However, this added flexibility introduces new challenges in circuit compilation.
Inspired by the placement and routing strategies for FPGAs, we propose to map all data qubits to fixed atoms while utilizing movable atoms to route for 2-qubit gates between data qubits.
Coined \textit{flying ancillas}, these mobile atoms function as ancilla qubits, dynamically generated and recycled during execution. 
We present Q-Pilot, a scalable compiler for FPQA employing flying ancillas to maximize circuit parallelism.
For two important quantum applications, quantum simulation and the Quantum Approximate Optimization Algorithm (QAOA), we devise domain-specific routing strategies.
In comparison to alternative technologies such as superconducting devices or fixed atom arrays, Q-Pilot effectively harnesses the flexibility of FPQA, achieving reductions of 1.4$\times$, 27.7$\times$, and 6.3$\times$ in circuit depth for 100-qubit random, quantum simulation, and QAOA circuits, respectively.

\end{abstract}

\maketitle

\thispagestyle{firstpage}
\section{Introduction}
\begin{figure}[t]
    \centering
    \includegraphics[scale=.45]{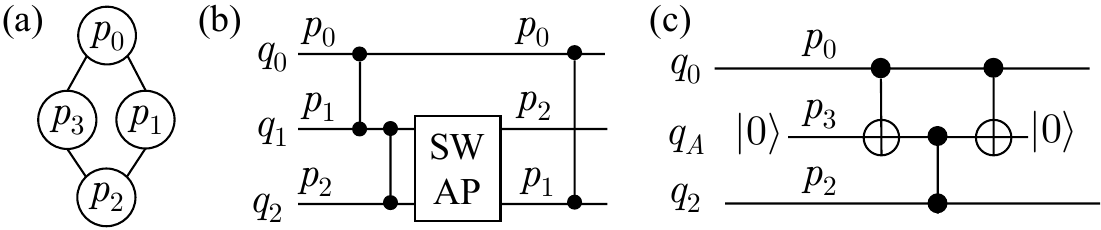}
    \vspace{-10pt}
    \caption{(a) The coupling graph of a QPU.
    (b) Qubit mapping and routing.
    The initial mapping is annotated at the beginning of each wire/qubit.
    A SWAP gate changes the mapping.
    (c) Using an ancilla and two more \cnot s to implement $\cz(q_0,q_2)$.}
    \label{fig:qubit-mapping-routing}
    \vspace{-10pt}
\end{figure}

\begin{figure}[t]
    \centering
    \includegraphics[width=\columnwidth]{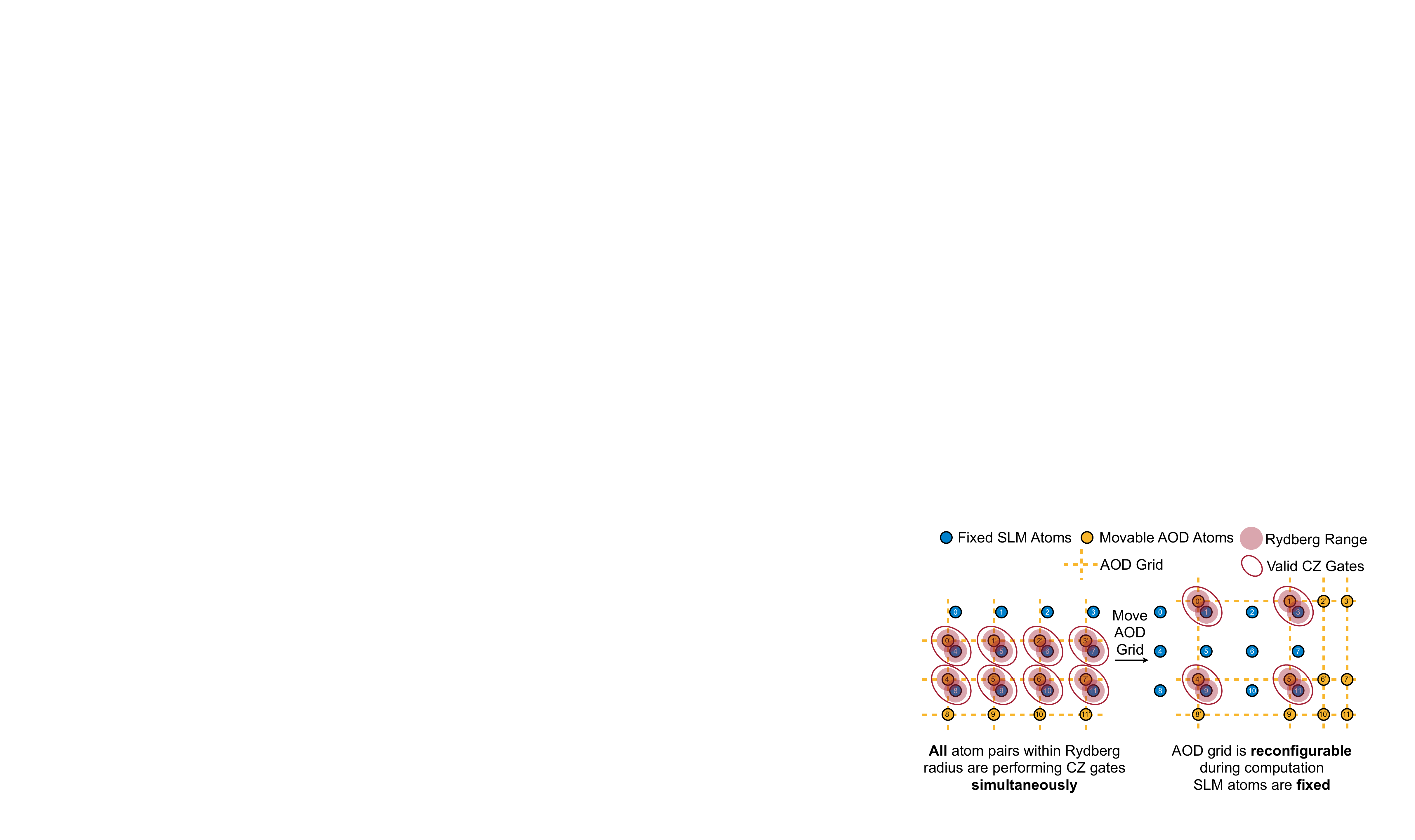}
    \caption{\fpqafull (\fpqa).}
    \label{fig:teaser}
\end{figure}

Quantum computing (QC)\thanks{Equal Contributions} hardware has seen rapid scaling, with superconducting systems offering up to 433 qubits~\cite{ibm433, rigetti, google72, intel49}, and neutral atom arrays reaching 1000+ qubits~\cite{quera256, atomcomputing}.
Utilizing these machines requires mapping qubits in a quantum program/circuit to physical qubits on the QPU, typically constrained by limited connectivity given by a coupling graph.
For example, Fig.~\ref{fig:qubit-mapping-routing}a illustrates a simple QPU with four physical qubits connected in a ring.
2-Q entangling gates, crucial for quantum programs, are restricted to adjacent physical qubits (e.g., $(p_0,p_1)$).
Consider a quantum program with gates $\cz(q_0, q_1)$, $\cz(q_1,q_2)$, and $\cz(q_2,q_0)$.
In Fig.~\ref{fig:qubit-mapping-routing}b, the initial qubit mapping is $q_i \mapsto p_i$ for $i=0,1,2$.
While this mapping supports the first two gates, $\cz(q_2,q_0)$ involves non-adjacent $p_2$ and $p_0$.
Here, a \swap\xspace gate is inserted to \textit{route} qubits, transforming the mapping.
However, \swap\xspace is costly: it can increase circuit depth, leading to more decoherence noise, and typically requires three 2-Q entangling gates, accumulating gate errors.
Given the current QPUs' relatively high noise levels, as quantum circuits grow, it becomes crucial that compilers minimize the overheads incurred by mapping and routing to optimize performance~\cite{asplos19-li-ding-xie-sabre-mapping, dac19-wille-burgholzer-zulehner-mapping-minimal-swaph, isca19-murali-linke-martonisi-abhari-nguyen-alderete-triq-architecture-studies, iccad21-tan-cong-qubit-mapping-absorption, iccad20-tan-cong-optimal-layout-synthesis, tcad08-maslov-falconer-mosca-placement, cgo18-siraichi-santos-collange-pereira-qubit-allocation, date18-zulehner-paler-wille-efficient-mapping-ibmqx, zhou_monte_2020, DBLP:conf/micro/MolaviXDPTA22, fan-reinforcement-learning, alam2020noise, alam2020efficient, park22dac, wu22iccad}.

A recent breakthrough enables atom movement during quantum circuit execution~\cite{bluvstein2022quantum}, profoundly impacting compilation by introducing dynamic coupling graphs for QPUs, as opposed to static configurations (Fig.\ref{fig:qubit-mapping-routing}).
In this work, we focus on a field programmable qubit array (FPQA) architecture that incorporates this technology.
FPQA features two atom types (Fig.\ref{fig:teaser}): SLM atoms (blue) are \textit{fixed} atoms in traps generated by a spatial light modulator (SLM); AOD atoms (yellow) are \textit{movable} atoms in traps generated by a 2D acousto-optic deflector (AOD).
The 2D AOD, a product of two 1D AODs, allows us to specify $X$ coordinates for columns (yellow dashes) and $Y$ coordinates for rows.
Consequently, AOD qubits move \textit{by entire rows and columns}.
To avoid non-deterministic behavior from trap overlap, we \textit{prohibit AOD rows/columns from moving over others}.
Physically, atom movement is a high-fidelity operation primarily constrained by coherence time: with only $0.1\%$ coherence time, an atom can traverse a region for $\sim 2,000$ qubits~\cite{bluvstein2022quantum}.
These movements are explicitly applied for 2-Q gates, which are induced by a \textit{global} Rydberg laser activating all atoms.
If two qubits are within the \textit{Rydberg radius} $r_b$, they become `coupled,' enabling the application of a \cz\xspace gate by the Rydberg laser.
Moving atoms between circuit stages couples different qubit pairs, resulting in a dynamic coupling graph.
To avoid unintended 2-Q gates, other atoms must be \textit{sufficiently separated} ($>2.5r_b$).
The global Rydberg laser requires less control and calibration, enhancing the \textit{scalability} of FPQA compared to prior works~\cite{Saffman_2016, graham_multi-qubit_2022} where the laser \textit{individually} address qubits for 2-Q gates.

We introduce \name, the first scalable FPQA router drawing inspiration from FPGA placement and routing.
Our approach, termed ``routing with flying ancillas,'' involves qubit mapping akin to cell placement and the use of movable ancilla qubits to bridge fixed atoms, similar to FPGA routing.
The advantages of flying ancillas include 1) high-parallelism circuit execution, 2) scalable compilation, and 3) no atom transfer required during computation.
% transfer only happens when preparing for atoms with no information.
To boost parallelism and thus reduce circuit depth, we implement a high-parallelism generic router, dynamically arranging AODs and scheduling 2-Q gates.
The router heuristically schedules as many parallel executable gates as possible in one laser stage up to AOD movement constraints.
We also devise \textit{application-specific} strategies: for each Pauli string in quantum simulation, we create multiple ancillas for a ``root'' qubit and employ graph algorithm to find the longest chain in the SLM array to perform the gates; for QAOA, we create ancilla per qubit instead of per gate, and leverage the commutation of gates to maximize parallel execution and reduce depth.
Extensive experiments demonstrate that our \raa compilation framework outperforms the best baseline, achieving 1.4$\times$, 27.7$\times$, and 6.7$\times$ smaller circuit depth for 100-qubit random, quantum simulation, and QAOA circuits.

\section{Flying Ancillas} \label{sec:routing-ancillas}
\subsection{Motivating Example of Routing $\cz$}
Revisiting the issue of the last gate in Fig.~\ref{fig:qubit-mapping-routing}, (c) introduces an alternative using ancilla qubit $q_A$ at $p_3$ instead of the \swap: $q_A$ is initialized to $|0\rangle$, the three-qubit initial state (in order $q_0q_Aq_2$) can be written as $a|000\rangle+b|001\rangle+c|100\rangle+d|101\rangle$.
After the first $\cx$, it becomes $a|000\rangle+b|001\rangle+c|110\rangle+d|111\rangle$.
After the $\cz$, it becomes $a|000\rangle+b|001\rangle+c|110\rangle-d|111\rangle$.
After the second $\cx$, it is $a|000\rangle+b|001\rangle+c|100\rangle-d|101\rangle$, which is the same as the case where $\cz(q_0,q_2)$ acts on the initial state.
In this process, $q_A$ acts as a `fan-out' of $q_0$.
However, note that it is only on the Z basis. Hence, this method's effectiveness hinges on the targeted 2-Q gate, specifically \cz\xspace in our case (and \zz\xspace later on), but it is not universally applicable.
Thus, we decompose other 2-Q gates using \cz\xspace or \zz\xspace beforehand.
Some previous works~\cite{fanout05, gokhale-qce21-fanout} leveraged these fan-outs to reduce cost in circuit synthesis, but we apply them in routing because uniquely in FPQA, the fan-out qubits can move physically.
If we rely on \swap s for the routing, the depth increases by 3 because we need 3 \cnot \xspace for 1 \swap, yet the new approach only increases depth by 2.

\subsection{General Theory of Routing $\cz$s with Ancillas}
\begin{figure}[t]
    \centering
    \includegraphics[scale=.45]{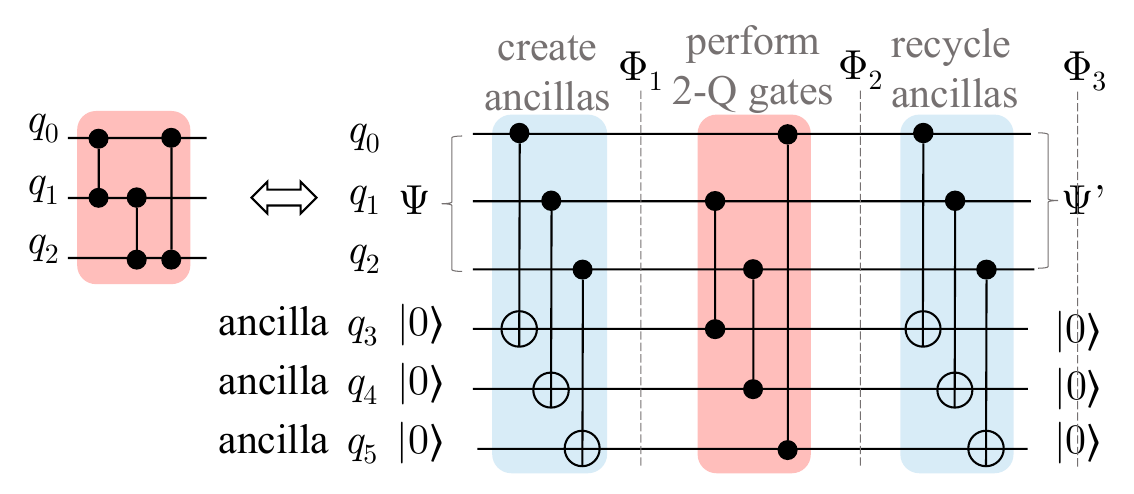}
    \caption{The general case of routing \cz s with ancillas.
    The 3 \cz s on the right can be executed simultaneously.}
    \label{fig:ancilla-harder-example}
\end{figure}

We prove a general result independent of the coupling graph.
Given an arbitrary $n$-qubit state $\Psi = C_0 |0\rangle + C_1|1\rangle + ... + C_{2^n-1}|2^n-1\rangle$, and a set of qubit pairs $\mathcal{C}$, applying $\cz_{j,j'} \ \forall (j,j')\in \mathcal{C}$ yields
\begin{equation} \label{eq:cz-effect}
    \Psi' = \left(\prod_{(j,j')\in \mathcal{C} } \cz_{j,j'}\right) \Psi  
    = \sum_{x=0}^{2^{n}-1} C_x\prod_{(j,j')\in \mathcal{C} } (-1)^{x_j x_{j'}} |x\rangle,
\end{equation}
where $x_j$ is the $j$-th bit of $x$.
We consider an alternative procedure as illustrated in Fig.~\ref{fig:ancilla-harder-example} where we 1) apply transversal $\cx$s from the $n$ qubits to $n$ fresh ancillas yielding $\Phi_1$, 2) apply one of the four possibilities (2 choices of whether to $+n$ for 2 indices) $\cz_{j(+n),j'(+n)}\ \forall (j,j')\in\mathcal{C}$ yielding $\Phi_2$, and 3) apply transversal $\cx$s again yielding $\Phi_3$.
We prove that $\Phi_3=\Psi'\otimes |0^n\rangle$, so our procedure is equivalent to applying the original $\cz$s in Eq.~\ref{eq:cz-effect}.

For every basis state $|x\rangle$ appended with $n$ fresh ancillas, applying transversal $\cx$s flips the ancilla state to $|x\rangle$.
Thus,
\begin{equation} \label{eq:transversal-first}
    \Phi_1 = \left(\prod_{i=0}^{n-1} \cx_{i,i+n}\right) \left(\Psi\otimes |0^n\rangle \right) 
    =\sum_{x=0}^{2^{n}-1} C_x|\overline{xx}\rangle,
\end{equation}
where an overhead line denotes the concatenation of bit-strings.
Then, for every pair $(j,j')\in\mathcal{C}$, we apply one of the 4 possible $\cz$s,
\begin{equation} \label{eq:cz-on-ancilla}
\begin{split}
    \Phi_2\ & = \left(\prod_{(j,j')\in \mathcal{C} } \cz_{j(+n),j'(+n)}\right) \Phi_1
     = \sum_{x=0}^{2^{n}-1} C_x
     \prod_{(j,j')\in \mathcal{C} } \bigg( \\ & (-1)^{\overline{xx}_{j(+n)} \cdot \overline{xx}_{j'(+n)}} \ |\overline{xx}\rangle \bigg)
    = \sum_{x=0}^{2^{n}-1} C_x\prod_{(j,j')\in \mathcal{C} } (-1)^{x_j x_{j'}} |\overline{xx}\rangle,
\end{split}
\end{equation}
where we use the fact that both the $j$-th bit (from the left) and the $(j+n)$-th bit of $\overline{xx}$ equals the $j$-th bit of $x$, similarly for $j'$.
Applying transversal $\cx$s, again, flips every state $|\overline{xx}\rangle$ back to $|x\rangle|0^n\rangle$, i.e.,
\begin{equation} \label{eq:transversal-second}
    \Phi_3 = \left(\prod_{i=0}^{n-1}\cx_{i,i+n}\right) \Phi_2 = \Psi'\otimes|0^n\rangle,
\end{equation}
which finishes our proof.

Note that $\cz$ gates are commutable, so the ones in Eq.~\ref{eq:cz-on-ancilla} can be applied in any order, which may unlock some freedom in scheduling.
Moreover, for each $(j,j')\in\mathcal{C}$, we have 4 possible $\cz$s to use in Eq.~\ref{eq:cz-on-ancilla} and many of them can be parallelized.
For example, in Fig.~\ref{fig:ancilla-harder-example}, $n=3$, and the original $\cz$s are $\mathcal{C}=\{(0,1), (1,2), (2,0)\}$ which takes at least 3 steps.
Using the procedure just presented, the $\cz$s on $(0+n,1)$, $(1+n,2)$, $(2+n,0)$ can be scheduled to just one step.

\subsection{Flying Ancillas in FPQA}

The flying ancillas scheme proves particularly advantageous for FPQA over other QC platforms, owing to its high-fidelity movements.
The most similar setting is in a multi-chain ion trap QPU~\cite{ionq-multi-chain}, where chains of ions are laid out in 1D, and two chains can be moved to merge or split again.
However, because there is no distinction between stationary and movable qubits like in FPQA, moving a regular qubit in the ion trap quantum computer has the same cost as moving an ancilla, so the flying ancilla scheme does not hold a big advantage.
Additionally, the limited number of qubits available on ion trap QPUs discourages leveraging numerous ancillas. Flying qubits, typically optical, are also employed as communication resources between individual superconducting QPUs but face challenges, including a low interfacing fidelity of approximately 80\% per flying qubit~\cite{microwavelink}.
In contrast, in FPQA, the two extra \cnot s required by flying ancilla can achieve 99.5\% fidelity, and the ancilla movement has negligible error~\cite{evered2023highfidelity}.
Despite this high fidelity, the state-of-the-art FPQA compilation work~\cite{tan2022} primarily utilizes only the movement of data qubits for routing, overlooking the potential advantages of routing via flying ancillas.

\section{Routing Framework}
\begin{figure*}[t]
    \centering
    \includegraphics[width=\textwidth]{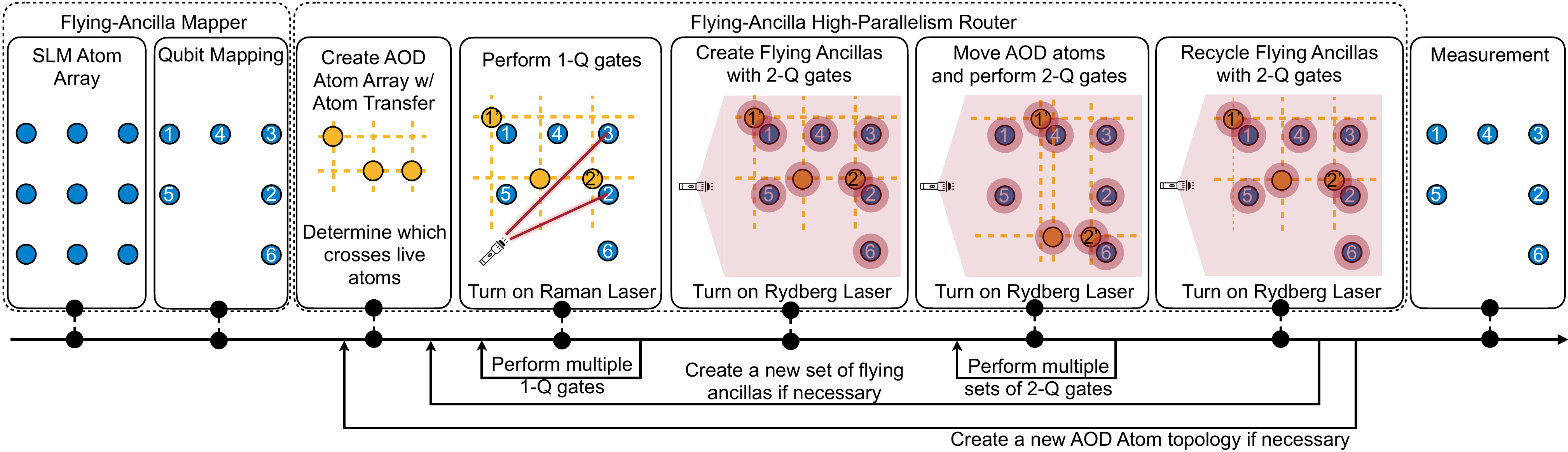}
    \caption{The flowchart of the \fpqa compilation framework. }
    \label{fig:compilation_flow}
\end{figure*}

\subsection{Overview}
% Our proposed iterative compilation process for \fpqa is shown in Fig.~\ref{fig:optimization_overview}. 
Given a target problem, the input values to the router are (1) the SLM array parameters (\#rows, \#columns, and locations), (2) the AOD array parameters (\#rows and \#columns), and (3) the qubit mapping.
We focus on routing in this work, so we simply map qubits in reading order throughout.
We refer to them as \emph{configurations} of the \fpqa.
Based on a configuration and the target problem, we leverage a high-parallelism router to generate an optimized schedule.
A fast performance evaluator can efficiently return the corresponding performance metric or cost, including the number of 1-Q gates, 2-Q gates, the circuit depth, and movement distance, which are closely related to the circuit fidelity.

With this performance evaluator, our compiler also supports router-in-the-loop \fpqa architecture design space exploration.
We can use the evaluated cost as feedback to optimize a configuration that targets higher circuit fidelity iteratively. 
After certain epochs, the compiler will output the best configuration and optimized schedule.
\subsection{Compilation of General Quantum Circuit}

The compilation process is shown in Fig.~\ref{fig:compilation_flow}. Given a quantum circuit, we first decompose the target circuit into 1-Q rotations and 2-Q \cz\xspace gates. Then, the gates are performed in alternating 1-Q and 2-Q stages. In the 1-Q stages, we turn on the individual addressable Raman laser to perform the desired gates on the target qubits.
After all the available 1-Q gates are done, we move to 2-Q stages. In such stages, we select a set of \cz~gates from the non-dependent front-layer of the circuit that can be performed in parallel. Then, we create flying ancillas from the control qubits and move the ancillas close to their corresponding target qubits. We turn on the Rydberg laser so that the ancillas will perform \cz~gates with the target qubits.
At last, we recycle these ancillas with \cx~gates and repeat this process. After all gates are done, we perform measurements and get results.

\begin{algorithm}[t]
    \SetAlgoLined
    \SetKwInOut{Input}{Input}
    \SetKwInOut{Output}{Output}
    \Input{Gates $G$ in a quantum circuit}
    $s\gets\emptyset$;~\text{Initial schedule}\;
    $g\gets G$;~\text{Initial gate candidates}\;
    $i\gets 0$\;
    \While{$g\neq\emptyset$}{
        $p_i=\texttt{FrontLayer}(g)$\;
        $q_i=\emptyset$\;
        \For{$p \in p_i$}{
         \If{$\texttt{IsLegal}(q_i \cup p)$}{
            $q_i\gets q_i\cup p$\;
            $g\gets g~\backslash~p$\;
         }
        }
        $s \gets s \cup \texttt{GenerateSchedule}(q_i)$\;
        $i \gets i + 1$\;
    }
    \Output{Schedule $s$ with maximum depth $i$}
    \caption{Generic router for arbitrary quantum circuits}
    \label{alg:alg_generic_router}
\end{algorithm}
    % \vspace{-5pt}

We first introduce the \emph{generic router} for arbitrary quantum circuits. We first transpile a given circuit with the
universal gate set: CZ+1Q gates. Then, we iteratively extract the front layer of that circuit. If the front layer contains 1Q gates, we perform the 1Q gates first. After that, the front layer contains only a set of CZ gates.

\begin{figure*}[t]
    \centering
    \includegraphics[width=\textwidth]{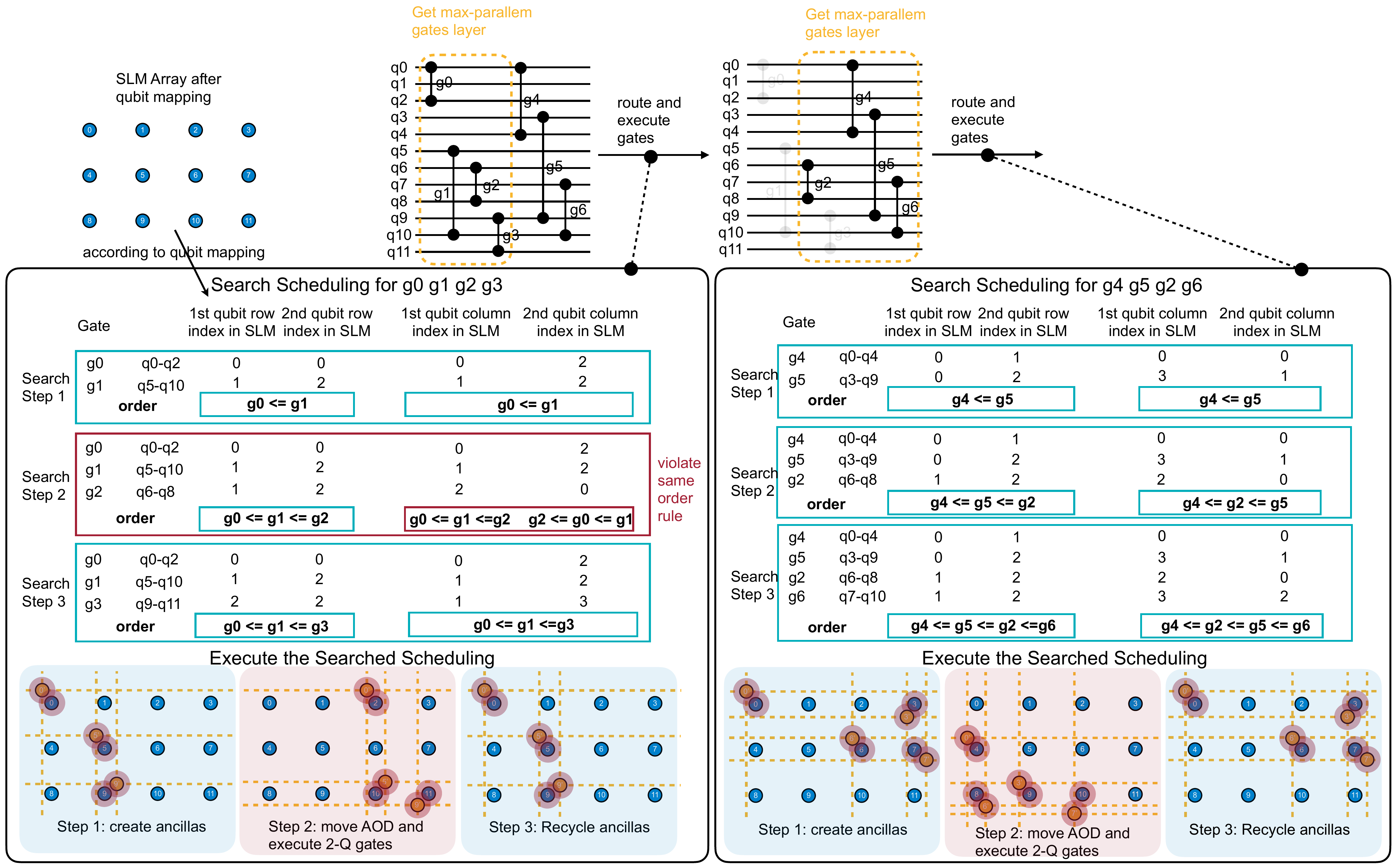}
    % \vspace{-20pt}
    \caption{Routing process of an example circuit using the generic router. The router adds as many as possible to one stage to execute them simultaneously.}
    \label{fig:general_router}
    % \vspace{-10pt}
\end{figure*}

To maximize the parallelism of the generated schedule while maintaining good scalability, we propose a heuristic-based scheduler that selects as many \cz~gates as possible while honoring the constraint of \fpqa in a single 2-Q gate stage described in Alg.~\ref{alg:alg_generic_router}. 
We show an example in Fig.~\ref{fig:general_router}.
Given a quantum circuit, we first detect the \emph{source layer of the dependency graph} according to the gate dependency, defined as the maximum potentially parallelizable gates, e.g.,  $(g_0,g_1,g_2,g_3)$. 
From the $i$-th source layer, we use a greedy algorithm to decide (1) the maximum legal subset of the source layer gates and (2) the grid locations of the AOD ancilla qubits.
The key intuition behind this greedy heuristic is that the order of rows and columns cannot be reversed.
We first sort the candidate gates by the index of the first qubit in the gate. In the first search step, the router adds $g_0$ and $g_1$ to the subset and checks the legality according to the qubit ordering in rows and columns.
Since the row/column order of $g_0$ and $g_1$ all satisfies $g_0<=g_1$, this is confirmed to be a legal subset.
Then, in the next step, the router tries to add $g_2$ to the subset.
However, in the column dimension, the 1st qubit is of order $g_0<=g_1<=g_2$, which conflicts with the 2nd qubit order $g_2<=g_0<=g_1$.
In other words, it is impossible to move AOD ancilla qubits to enable parallel execution of those three gates.
This violation of the order rule will kick gate $g_2$ out of the legal subset.
Similarly, the router adds $g_3$ into the subset and ultimately finds the maximum legal subset $(g_0,g_1,g_3)$ and the corresponding grid locations of three AOD ancilla qubits, i.e., (0,0), (1,1), and (2,2).
This is the end of the sub-schedule for the $i$-th stage.
To physically execute this $i$-th sub-schedule, we perform a three-phase operation: (1) move the AOD ancilla qubits near the target SLM qubits and \emph{copy} their states via \cx s; (2) move the ancilla qubits to the target qubits to parallel execute the gates in the legal subset; and (3) finally recycle the states from ancilla qubits to the original SLM qubits via \cx s. After performing a set of 2-Q gates, the router will repeat the process on the next \textit{source layer} of the circuit $g_4, g_5, g_2, g_6$. The router will repeat this legal subset selection process on the remaining gates until all gates are executed in the schedule and ultimately output the complete schedule. 

Note that the AOD configuration between two iterations can be different. For our example, the ancillas live on different crosses. This can be achieved by transferring a set of spare SLM atoms to arbitrary rows and columns on the AOD grid. The atom loss is not fatal in our scheme because they do not contain any quantum information during transfer.

% \begin{figure}[t]
\begin{algorithm}[t]
    % \SetAlgoLined
    \SetKwInOut{Input}{Input}
    \SetKwInOut{Output}{Output}
    \KwData{List $P$: qubits in the Pauli string with non-$I$ paulis}
    $s \gets \emptyset$; \text{Schedule for the compiled program}\\ 
    $P \gets P~\backslash~P[0]$; \text{$P[0]$ is the root qubit}\\
    % $\text{AOD ancilla qubit preparation: copy state of}~q_i$\;
    $g\gets \emptyset$;~\text{Directed compatibility graph. Two qubits }\\\text{are compatible if and only if there is a path between them.}\\
    \For{$q_i\in P$} {
             $g.nodes \gets g.nodes \cup q_i$\; 
            
        }
        \For{$q_i\in P$} {
               \For{$q_j\in P~\backslash~q_i$} {
                  \If{$q_j.row >= q_i.row~\&~ q_j.col >= q_i.col$} { $g.edges\gets g.edges\cup (q_i,q_j)$\;}
            }
        }
    \While {$g \neq\emptyset$} {
        $\text{Find the longest path } l \text{ in }g$\;
        $s \gets s\cup \texttt{GenerateSchedule}(l)$;\\
        $g \gets g~\backslash~\{n|n\in l\}$;
    }
    \KwResult{Schedule $s$}
    \caption{Customized router for quantum simulation}
    \label{alg:alg_dp_simulation}
\end{algorithm}
% \end{figure}

\begin{figure}[t]
    \centering
    \includegraphics[width=0.9\columnwidth]{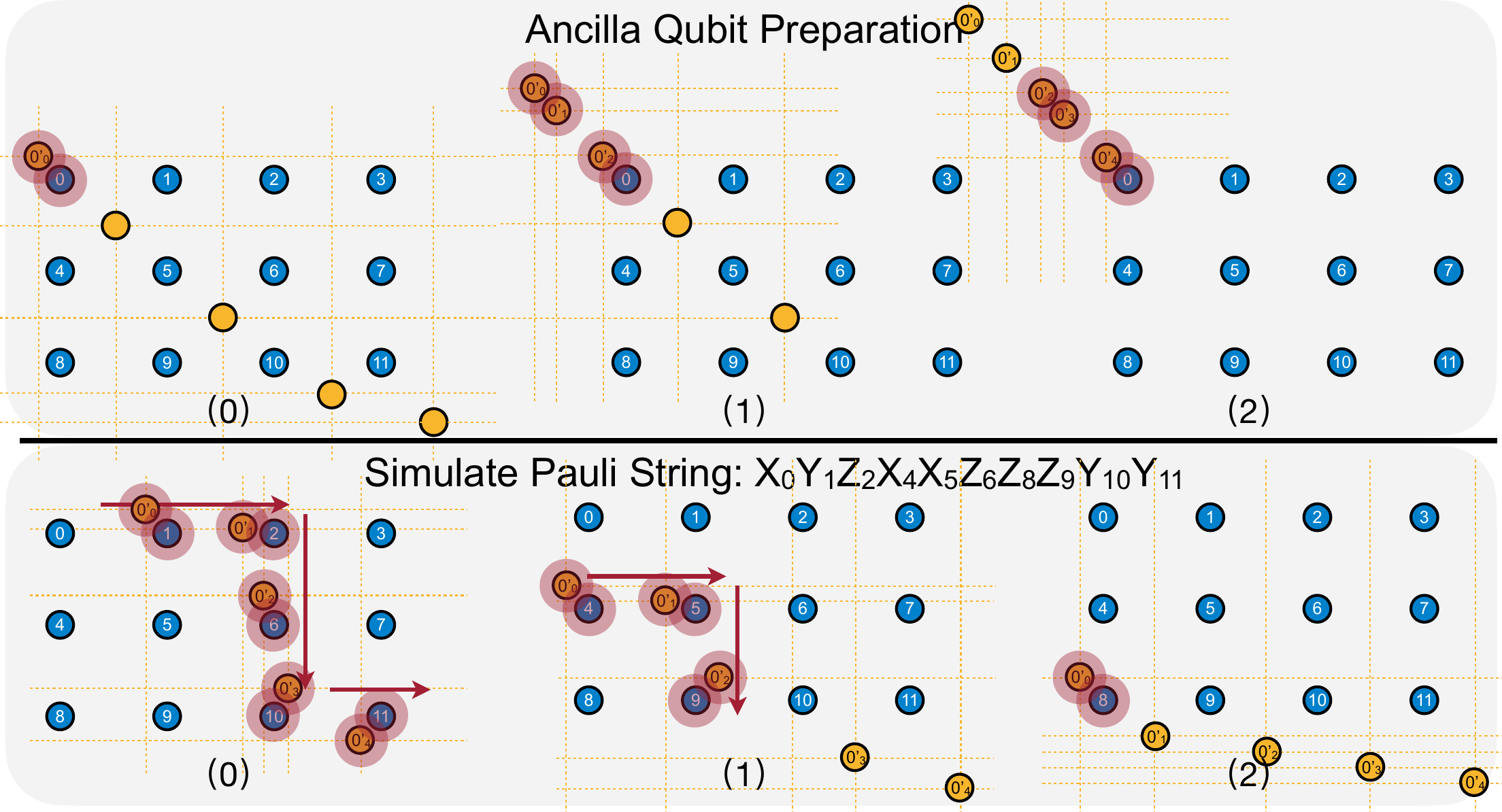}
    \caption{\name routing quantum simulation circuits. 
    % In the first plot, multiple flying ancillas are created for $q_0$ in a few stages. The second plot demonstrates the parallelized process for compiling the quantum simulation circuits corresponding to given Pauli strings.
    }
    \label{fig:q_sim_schedule0}
\end{figure}

\subsection{Customized Router for Quantum Simulation}
For specific applications, we propose domain-specific routing strategies for higher parallelism. The first application is quantum simulation. To simulate the evolution under a Pauli string, the core part of the simulation algorithm works as follows: First, select a starting qubit inside the given Pauli string and then perform \cx s on all pairs between the starting one and other qubits in the string.

We propose a longest path-based algorithm to compile this problem on the \fpqa, described in Alg.~\ref{alg:alg_dp_simulation}. We configure the AOD array so that all ancilla qubits are on the \textit{diagonal} of the grid and can be moved with the best flexibility. Then, we select the qubit $i$ with the smallest index and fan-out its state to all AOD ancilla qubits. To maximize the parallelism, we need to find the longest legal path in the directed dependency graph, where each qubit points to all other qubits in its lower-right corner, as shown in Fig.~\ref{fig:q_sim_schedule0}.

Given this longest path, we move the AOD ancilla qubits to their target SLM qubits and perform \cx s in parallel.
Further, those executed qubits will be removed from the candidate set, and the longest path-finding procedure will repeat until all gates are executed.
Note that this longest path-finding can be implemented efficiently with dynamic programming.
Compared to the generic router, which applies atom transfer to create and recycle ancilla qubits at each stage, this specialized quantum simulation router will maintain the states on the ancilla qubits across stages for one Pauli string, thus having a lower overhead.
To generate N fan-out qubits for N non-identity Pauli operators in a string, we initiate a fan-out operation by relocating the root SLM qubit near the target and executing a \mbox{\cx} gate, as indicated by the underline X below. Additional fan-out qubits can be generated by performing \cx operations with adjacent qubits. This leads to a geometric progression in the number of new fan-out qubits—1, 2, 4, 6, 8, etc.—yielding a circuit depth of $O(\sqrt{N})$.
\begin{table}[h]
\setlength{\tabcolsep}{1.2pt}
\centering
\[
\begin{array}{c@{}c}
\begin{tabular}{*{25}{c}}
O & O & O & O & O & O & O & O & O & O & O & O & O & O & O & O & O & O & O & O & O & O & O & O & \\
O & O & O & O & O & O & O & O & O & O & O & O & \underline{X} & O & O & O & O & O & O & O & O & O & O & O & +1 \\
O & O & O & O & O & \underline{X} & O & O & O & O & O & O & X & X & O & O & O & O & O & O & O & O & O & O & +2 \\
O & O & O & O & O & X & X & O & O & O & O & X & X & X & X & O & O & O & \underline{X} & O & O & O & O & O & +4 \\
O & O & O & O & X & X & X & X & O & O & X & X & X & X & X & X & O & O & X & X & O & \underline{X} & O & O & +6 \\
O & \underline{X} & O & X & X & X & X & X & X & X & X & X & X & X & X & X & X & X & X & X & X & X & X & O & +8 \\
X & X & X & X & X & X & X & X & X & X & X & X & X & X & X & X & X & X & X & X & X & X & X & X & \\
\end{tabular} & \rotatebox{270}{$\xrightarrow{\text{time}}$} \\
\end{array}
\]
\label{tab:fanout}
\end{table}

\subsection{Customized Router for QAOA}
\begin{algorithm}[t!]
\caption{QAOA compilation with flying ancilla and high-parallelism router}
\label{algo:qaoa}
\textbf{Input:}
edges: List[(q1,q2)] is a list of edges\\
Program: an empty quantum program\\
\While {edges is not empty}{
    cancel\_pairs\_first\_row=[]\\
    Find e0 in edges with smallest e0.q1.\\
    cancel\_pairs\_first\_row.append(e0)\\
    Program.cancel(e0)\\
    \While {True}{
        Find e in edges with smallest e[0], which is compatible and e.q1.y=e0.q1.y.\\
        \If{can't find}{
            Break
        }
        cancel\_pairs\_first\_row.append(e)\\
        Program.cancel(e)
    }
    \For{i = 1$\ldots$n\_row, i' = i$\ldots$n\_row}{
        match=True\\
        \For{e in cancel\_pairs\_first\_row}{
            replace e.q1.y with i, replace e.q2.y with i'\\
            \If {e not in edges}{
                match=False
            }
            \If{match}{
                \For{e in cancel\_pairs\_first\_row}{    
             replace e.q1.y with i, replace e.q2.y with i'\\
             Program.cancel(e)
             }
            }
        }
    }
}
\textbf{Output:} Compiled Program
\end{algorithm}

\begin{figure}[t]
    \centering
    \includegraphics[width=\columnwidth]{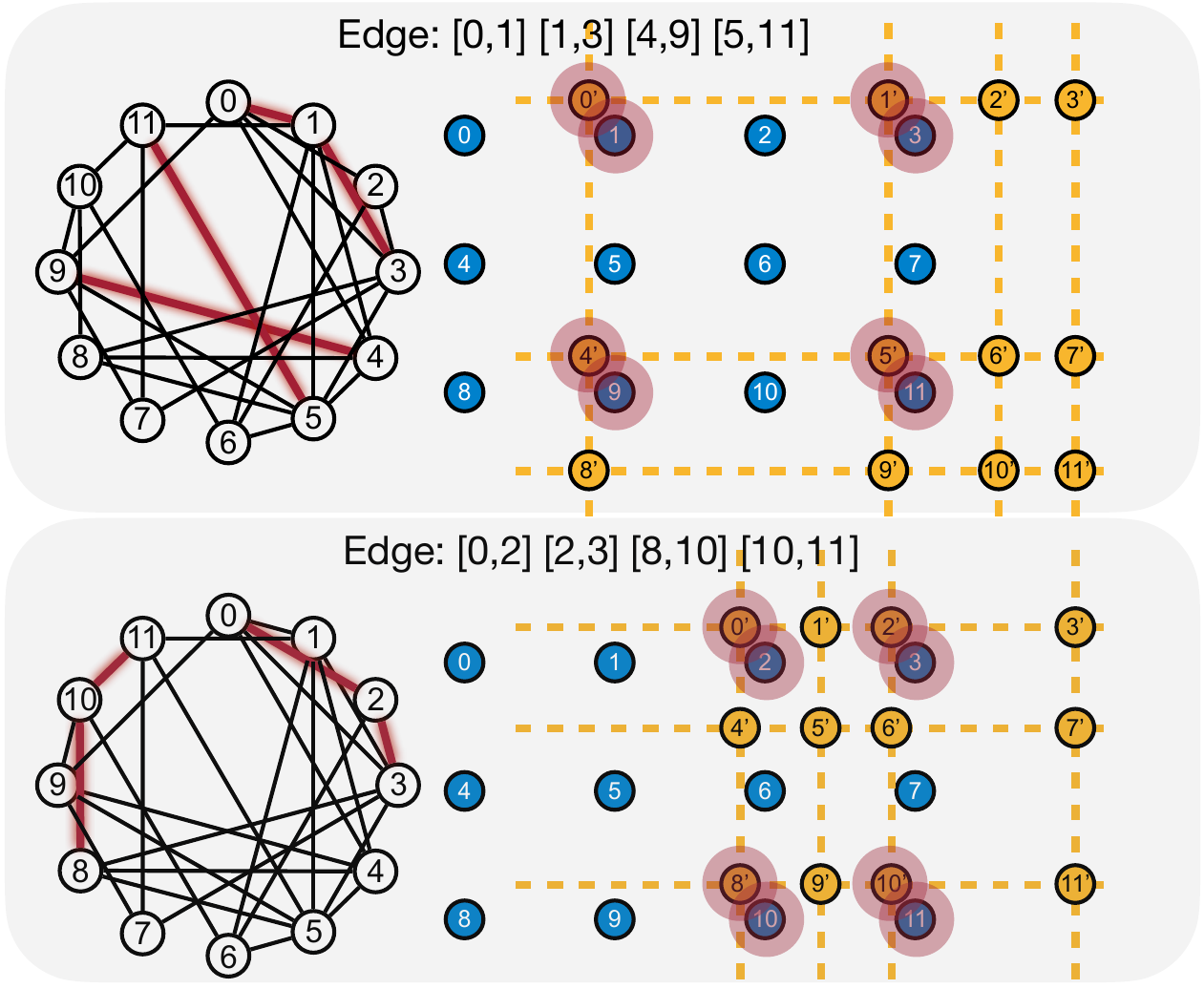}
    \vspace{-20pt}
    \caption{Scheduling of a QAOA circuit using \name. The graph representation of the circuit is shown on the left. 
    The edges correspond to interactions between two qubits.
    % The subgraphs are ordered in row-major order.
    % \SH{we should make the grid distance shorter}
    }
    \label{fig:qaoa_schedule0}
    \vspace{-10pt}
\end{figure}

Another task that can be highly parallel is Quantum Approximate Optimization Algorithm (QAOA). In QAOA, we are given a graph, and our target is to perform 2-Q gates on every edge in the graph shown in Fig.~\ref{fig:qaoa_schedule0}.

\begin{figure}[t]
    \centering
    \includegraphics[width=\columnwidth]{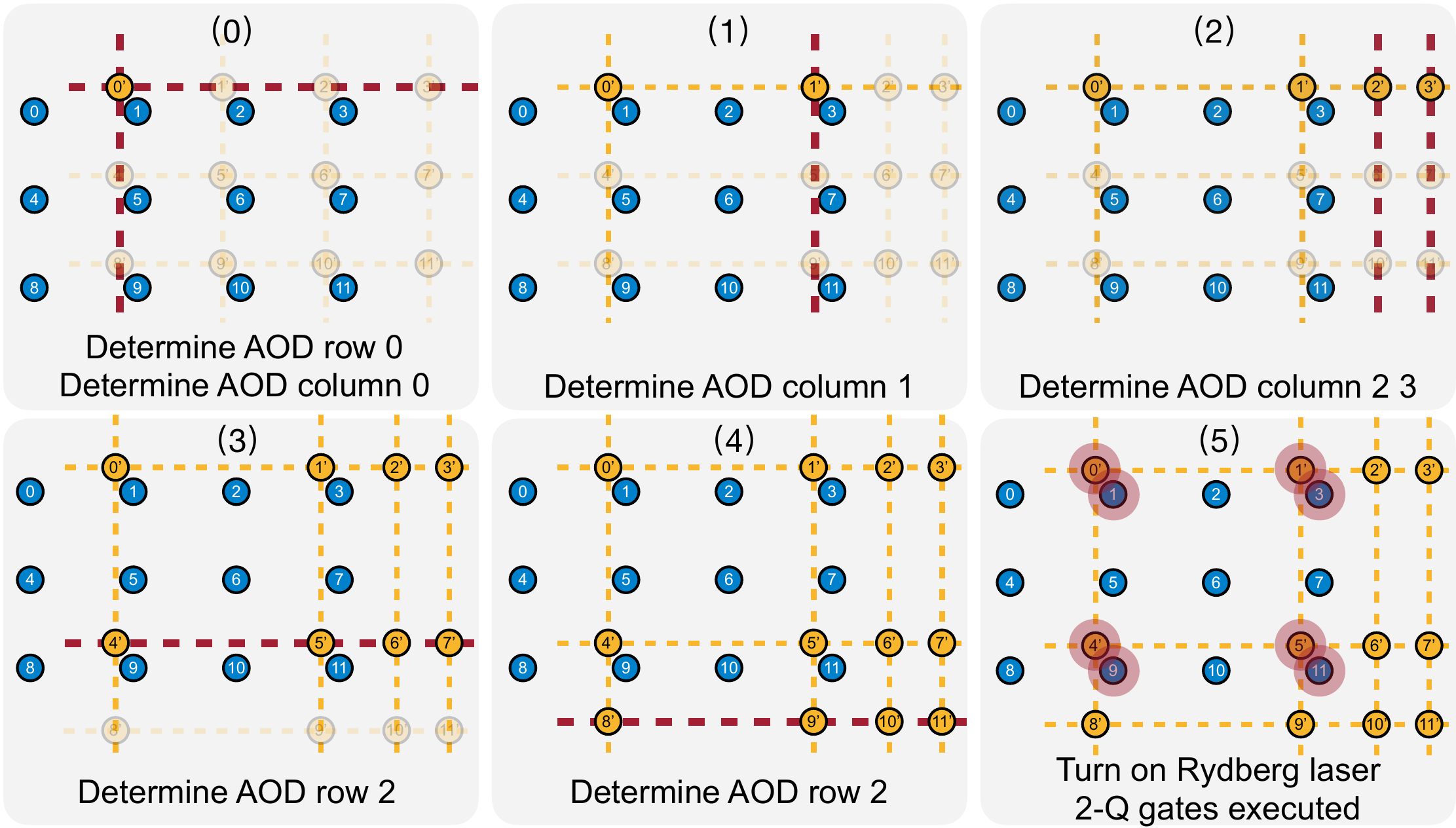}
    \caption{\name compiling one stage of an example QAOA circuit.}
    \label{fig:qaoa_one_stage}
\end{figure}

First, we create one ancilla for each qubit. These ancillas will be recycled once the whole graph is done. Then, our router completes this task in a multi-stage way. Each stage will perform one or multiple 2-Q gates corresponding to some edges in the graph.
We illustrate the detailed procedure of the first stage in Fig.~\ref{fig:qaoa_schedule0}. Among all the qubit pairs, we select the one with the smallest index as the highest-priority pair to begin with, e.g., (0',1).
Since each AOD row and column must move simultaneously, we check which pair can be performed inside the same row, e.g., (1',3) is matched in this case.
The rest of the AOD columns have already moved outside the SLM array.
Then, they will not interact with any SLM atoms.
Once the locations of all ancilla qubits in the first row are determined, other ancilla qubits on the rest of the rows can only move vertically due to the grid constraint.
Then, we need to determine the vertical location of each row one by one.
For the second AOD row, we found the best vertical location that can allow the most pairs to interact, e.g., in this case, we match two pairs (4',9) and (5',11). 
Note that any undesired interaction is illegal and thus should be avoided.
This process will repeat until no rows can legally interact with any SLM atoms.
Then, we can determine the locations of all AOD qubits and turn on the Rydberg laser to perform the parallel 2-Q gates.
This greedy algorithm, details in Alg.~\ref{algo:qaoa}, always tries to achieve the maximum matching on the first row and ultimately reaches a schedule with max parallelism.

So far, we have finished the first stage of the schedule with four 2-Q gates being performed. In the second stage, the highest-priority pair now becomes (0',2), and the same procedure as stage one can be applied to find a legal schedule with maximal parallelism.
After $t$ stages, the compilation flow ends with a $t$-stage legal schedule where all 2-Q gates are performed.
Lastly, we recycle the ancillas and complete the task.

\begin{figure}[t]
    \centering
    \includegraphics[width=\columnwidth]{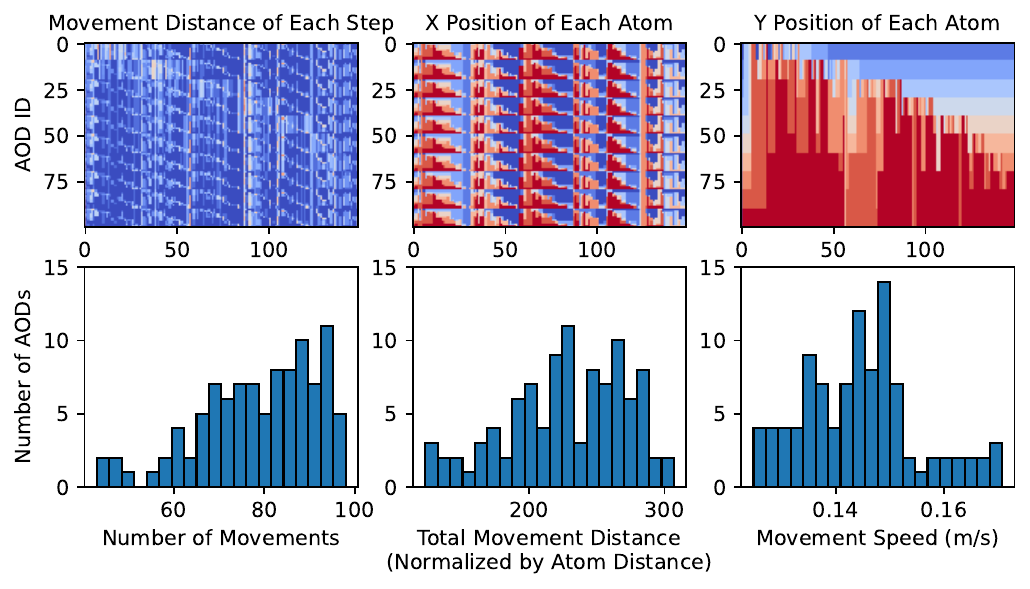}
    % \vspace{-15pt}
    \caption{Movement spatiotemporal patterns.
    }
    \label{fig:heatmap_qaoa}
    % \vspace{-10pt}
\end{figure}

\begin{figure}[t]
    \centering
    \includegraphics[width=\columnwidth]{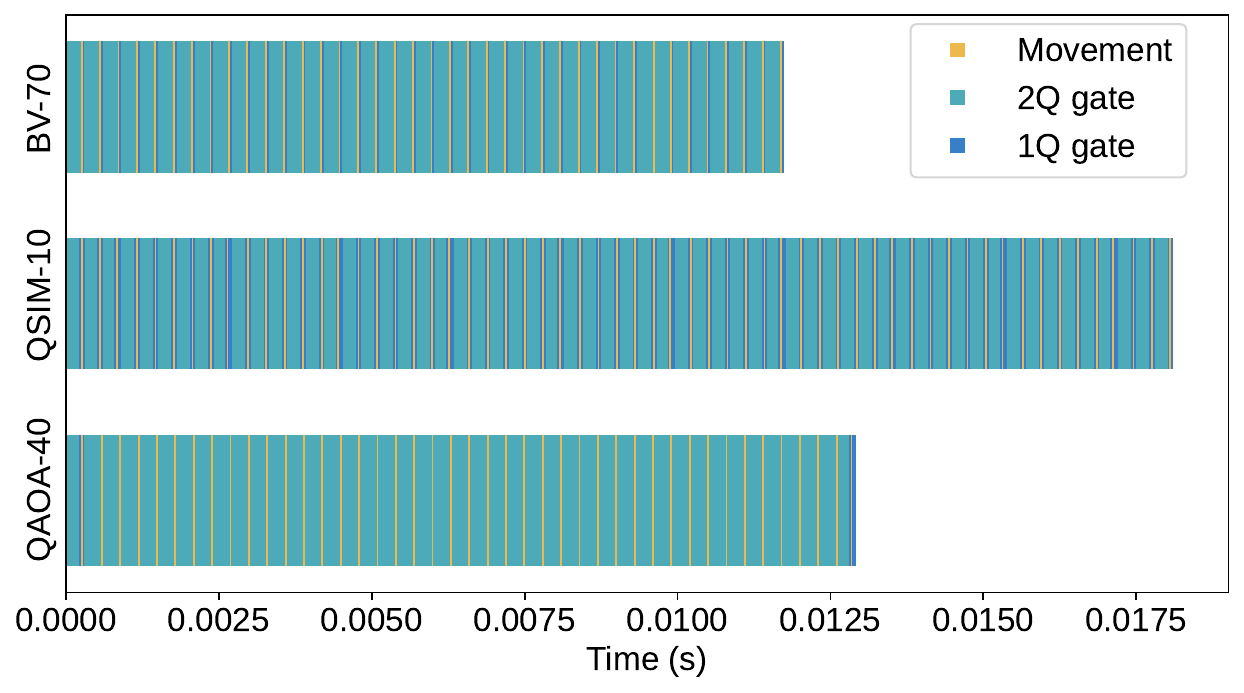}
    % \vspace{-5pt}
    \caption{
Detailed execution of compiled program.}
    \label{fig:trace}
    % \vspace{-10pt}
\end{figure}

Fig.\mbox{\ref{fig:heatmap_qaoa}} depicts the spatiotemporal dynamics of a 100-qubit QAOA circuit, revealing a periodic pattern in atom positions due to iterative optimization for parallel gate execution. The figure also includes histograms detailing the frequency of movements, distances traversed, and average speeds. Based on realistic parameters from~\mbox{\cite{bluvstein2022quantum}}, the typical speed is measured at 0.15 m/s.
Fig.~\mbox{\ref{fig:trace}} shows the execution timeline of a program compiled with the FPQA where movements are the largest part.

\section{Evaluation}

\subsection{Evaluation Methodology}

\textbf{Benchmarks.}
We utilize three benchmark types: random, quantum simulation, and QAOA circuits. 
Benchmarks were created for 5, 10, 20, 50, and 100 qubits. 
Random circuits were generated with Qiskit's \texttt{random\_circuit} function, which randomly places 1-Q and 2-Q gates on qubits. 
The number of \cx\ gates is set at 2x, 5x, 10x, 20x, and 50x the qubit count. 
Quantum simulation circuits were formed from 100 random Pauli strings. 
The probability $p$ of a qubit having a Pauli operator $X$, $Y$, or $Z$ varies from 0.1 to 0.5. 
QAOA circuits were constructed using $\zz$ gates between random qubit pairs. 
These pairs had an edge probability $p$ of 0.1 to 0.5. 
We also designed specific QAOA circuits based on 3-regular and 4-regular graphs. 
These circuits also used 5, 10, 20, 50, and 100 qubits.

\begin{figure}[t]
    \centering
    \includegraphics[width=\columnwidth]{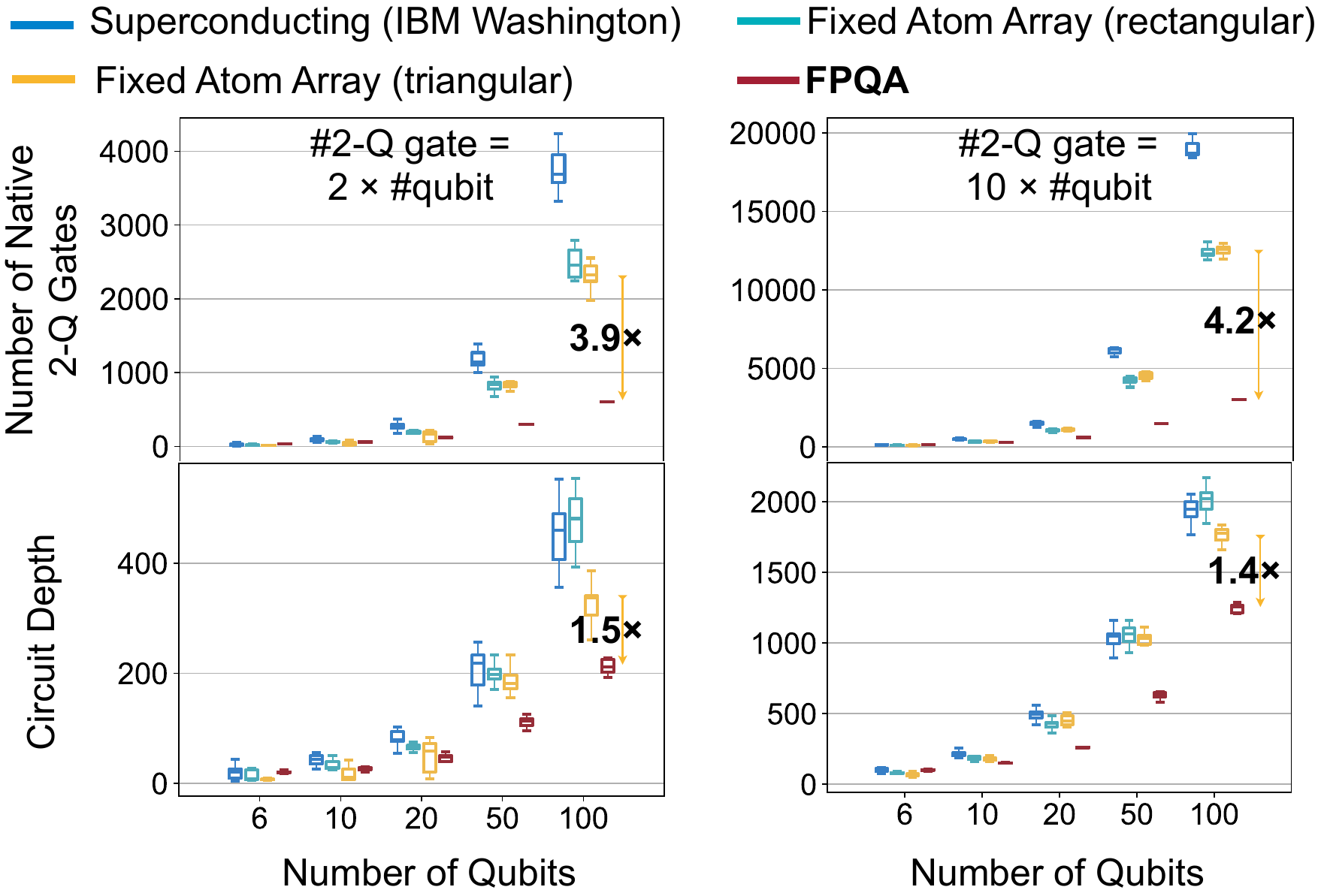}
    \caption{Comparison of compiled 2-Q gate count and circuit depth between \name and the three baselines on random circuits. The random circuits vary in size, from 5-Q to 100-Q, and have 2-Q gate count between 2$\times$ and 10$\times$ qubit count.
    }
    \label{fig:arb_main}
\end{figure}

\noindent\textbf{Baselines.}
3 devices as baselines were chosen: the 127-qubit IBM Washington machine, a 16$\times$16 square lattice, and a 16$\times$16 triangular lattice of fixed neutral atoms, following Ref.~\cite{tan2022}. The IBM machine features a heavy hexagon coupling graph. The square lattice's atoms connect to four nearest neighbors, while the triangular lattice's atoms connect to six. Qiskit's transpiler compiled benchmark circuits onto these devices at optimization level 3. Circuit depth, defined as the number of parallel 2-Q gate layers, was a key comparison metric, alongside the number of 2-Q gates in each \textit{compiled} circuit for the baseline devices and \name.
Additionally, \name was benchmarked against the solver-based compiler from Ref.~\cite{tan2022}, used for QAOA problems on 3- and 4-regular graphs. Comparisons included circuit depths and compilation times, with a 4,000s timeout ($\sim$an hour) set for the solver-based compiler due to its exponential runtime scaling.

\subsection{Main Results}
\noindent\textbf{Results on random circuits.} Fig.~\ref{fig:arb_main} shows the results of compiling random circuits. Compared with three baseline devices, for 100 qubits \name shows an average of 4.2$\times$ reduction in the compiled 2-Q gate count, as well as an average of 1.4$\times$ reduction in compiled circuit depth compared with the best-performing baseline approach.

\begin{figure}[t]
    \centering
    \includegraphics[width=\columnwidth]{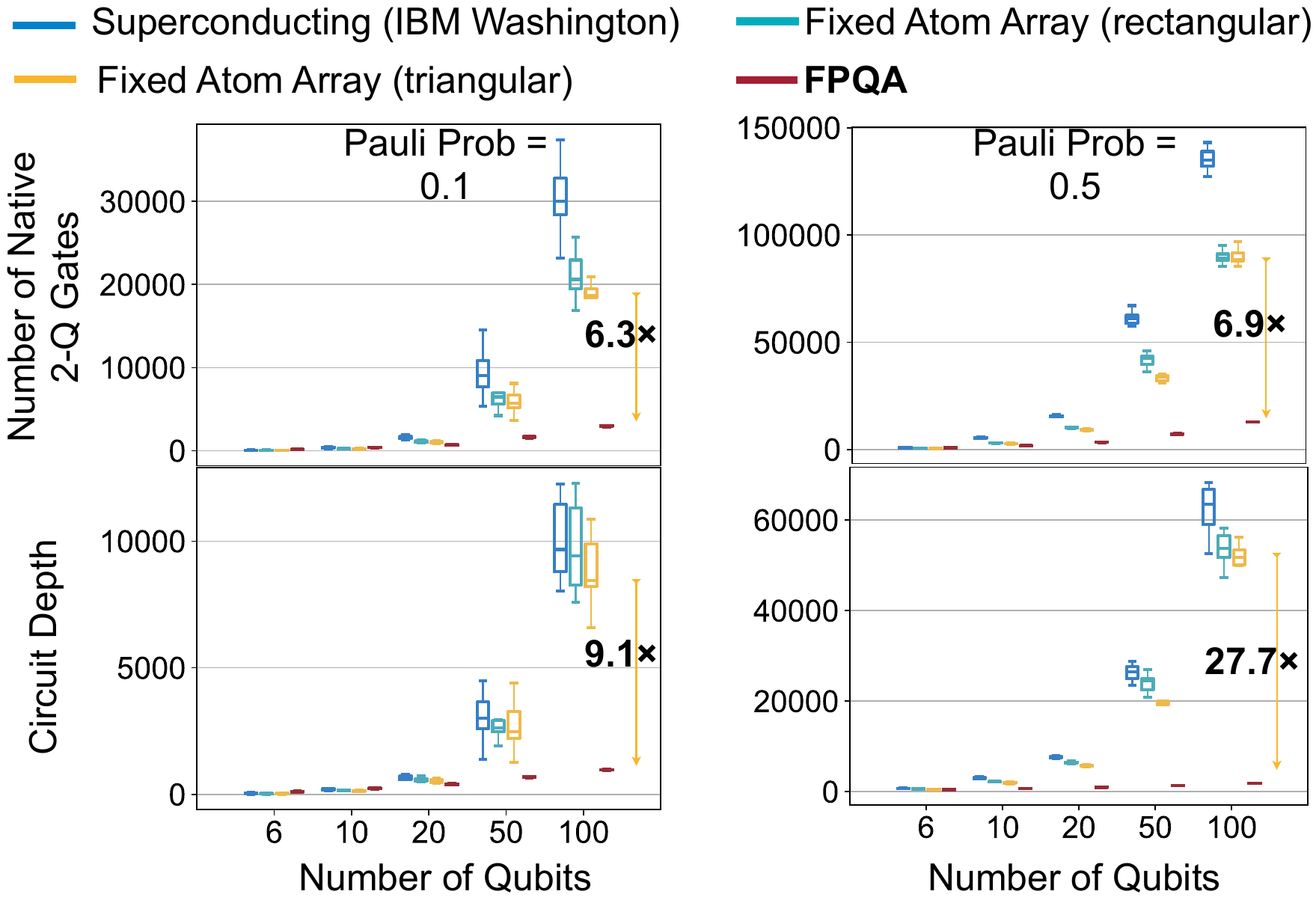}
    \caption{Comparison of compiled 2-Q gate count and circuit depth between \name and the three baselines on quantum simulation circuits from 5-Q to 100-Q. The circuits are generated with Pauli probability $p=0.1$ and $0.5$.
    }
    \label{fig:qsim_main}
\end{figure}

\noindent\textbf{Results on quantum simulation circuits.} Fig.~\ref{fig:qsim_main} shows the results of compiling quantum simulation circuits. For Pauli probabilities 50\%, \name shows an average of 6.9$\times$ reduction in the compiled 2-Q gate count and an average of 27.7$\times$ reduction in compiled circuit depth compared with the best-performing baseline on 100-qubit circuits compared with the best baseline. Besides the random Pauli strings, we also test with the Pauli strings used in some molecule simulation problems~\cite{li2022paulihedral}. As shown in Table~\ref{tab:qsim_fixed}, \name shows an average $1.36\times$ reduction in the 2-Q gate count and average $2.60\times$ circuit depth reduction over the best baseline.

\begin{figure}[t]
    \centering
    \includegraphics[width=\columnwidth]{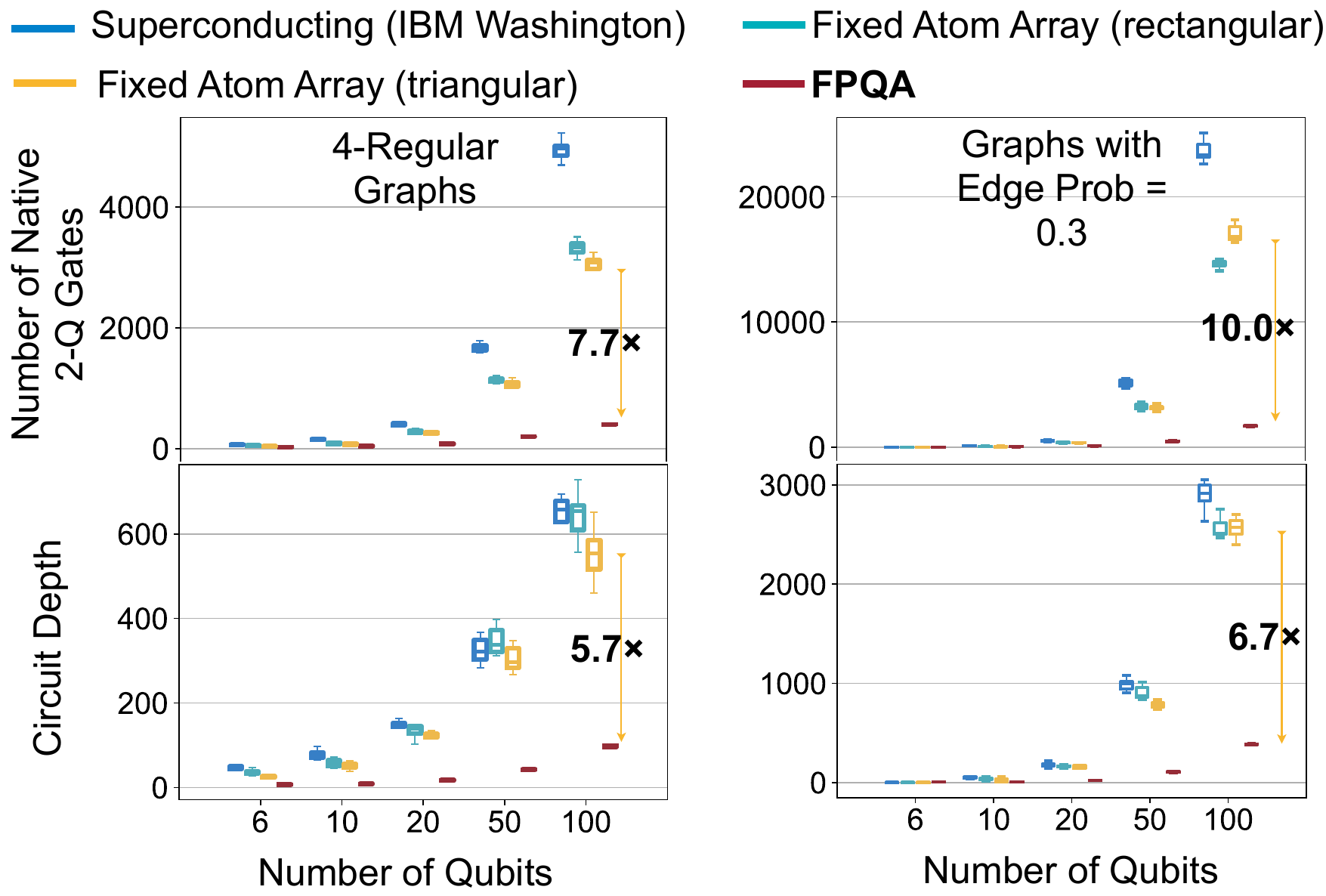}
    \caption{Comparison of compiled 2-Q gate count and circuit depth between \name and the three baselines on QAOA circuits. The QAOA circuits vary in size, from 6-Q to 100-Q, and are generated with edge $p=0.3$ and 4-regular graphs.
    }
    \label{fig:qaoa_main}
    % \vspace{-5pt}
\end{figure}

\noindent\textbf{Results on QAOA circuits.} Fig.~\ref{fig:qaoa_main} shows the results of compiling Max-Cut QAOA circuits for 4-regular graphs and random graphs with edge occupancy 30\%. \name again shows an average of 10.0$\times$ reduction in compiled 2-Q gate count and an average of 6.7$\times$ reduction in the compiled circuit depth.

\begin{table}[t]
\centering
\renewcommand*{\arraystretch}{1}
\setlength{\tabcolsep}{7pt}
\footnotesize
\caption{Quantum Simulation for Molecule Pauli strings. }
\vspace{-8pt}
\begin{tabular}{llcccc}
\toprule
Benchmark&Device & H2 & LiH\_UCCSD &H2O&BeH2 \\
\midrule
\multirow{4}{*}{Depth}&FAA(rectangular)&76&2,772&31,087&43,919\\
&FAA(triangular)&61&2,052&26,189&37,314\\
&Superconducting&77&3,403&40,080&59,259\\
&\textbf{Ours}&61&849&7,585&10,617\\
\midrule
\multirow{4}{*}{\#2Q Gate}&FAA(rectangular)&82&3,577&41,306&58,720\\
&FAA(triangular)&73&2,616&35,353&51,699\\
&Superconducting&85&5,082&67,247&103,594\\
&\textbf{Ours}&94&2,130&20,966&29,518\\
\bottomrule
\end{tabular}

% 0.00276
% 0.00774
% 0.0292
% 0.0785

\label{tab:qsim_fixed}
\vspace{-10pt}

\end{table}

\noindent\textbf{Comparison with the Solver-Based Compiler.}
As illustrated in Table~\ref{tab:comp_solver}, we compare \name against the solver-based methods~\cite{tan2022, tan2023compiling} in compiling QAOA circuits for regular graphs.
Ref.~\cite{tan2023compiling} relaxes the formulation of Ref.~\cite{tan2022} to tradeoff compilation time and quality.
While the these method achieve better solutions, they struggles with larger problems, often failing to find a solution within an hour due to its exponential runtime scaling.
In contrast, \name efficiently compiles all these problems in under 1 second, with the compiled circuit depth not exceeding 4$\times$ the optimal depth. 
%This performance highlights \ name's scalability and efficiency in handling large-scale problems.

\subsection{Analysis}

\begin{figure}[t]
    \centering
    \includegraphics[width=\columnwidth]{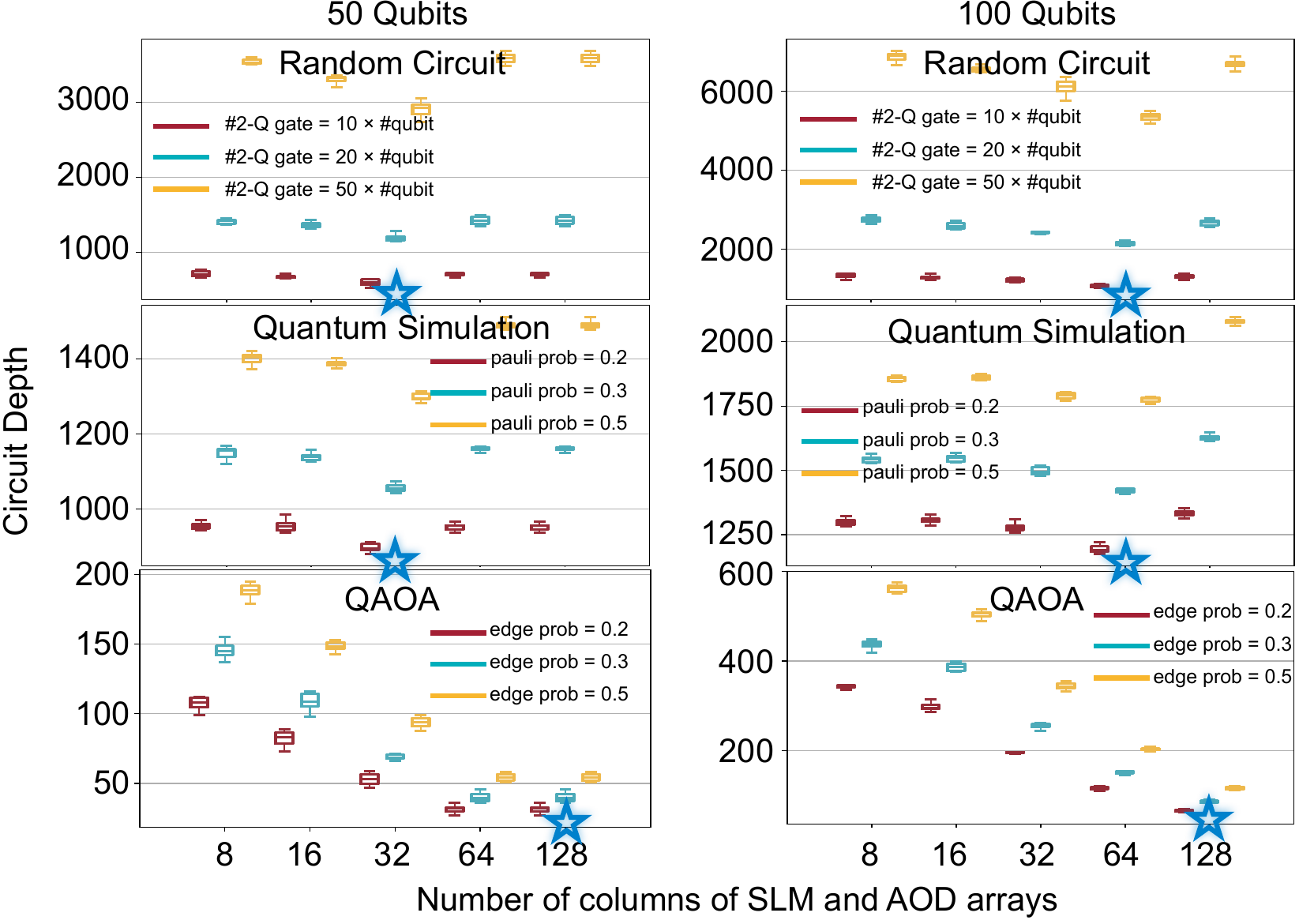}
    \caption{Circuit depth vs array width of SLM and AOD arrays in \fpqa. The three benchmark circuits are shown here for 50 qubits and 100 qubits. The star in each graph marks the optimal array width for the smallest circuit depth.}
    \label{fig:impact_array_size}
\end{figure}

\textbf{\textit{Impact of Array Size on Circuit Depth.}} Fig.~\ref{fig:impact_array_size} shows how array sizes affect the compiled circuit depth. We organized the qubits into rectangular arrays of varying widths ($8, 16, 32, 64, 128$), with the optimal array widths marked by stars in the figure. Optimal array widths vary across different problems, highlighting a tradeoff between greater parallelism within a row and across different rows. Specifically, in Fig.~\ref{fig:impact_array_size}, we observe that QAOA circuits achieve optimal performance with large array width (128), while random circuits and quantum simulation problems are best served with moderate array widths (64 or 32 in our study). The insight here is while larger array widths offer more parallel execution paths, they might not always correspond to increased efficiency for all types of problems, possibly due to overheads or specific characteristics of the circuit structure.

\begin{table}[t]
\centering
\renewcommand*{\arraystretch}{1}
\setlength{\tabcolsep}{2.8pt}
\footnotesize
\caption{Comparison of \name with solver based method.
% on Fashion-2 task in U3+CU3 space. 
% Under the same \#parameters, \name outperforms random and human designs. 
% Pruning reduces circuit depth and gate count, thus increasing measured accuracy.
}
\begin{tabular}{lccccccc}
\toprule
% Benchmark & Compilation time & Circuit Depth \\
% \midrule
% \multirow{2}{*}{3-regular 6Q}  & 0.1725 & 3\\
% % \midrule
% & 0.00131 & 6 \\
% \midrule
% \multirow{2}{*}{3-regular 10Q}  & 0.3811 & 3\\
% & 0.00276 & 7.7 \\
% \midrule
% \multirow{2}{*}{3-regular 20Q}  & 74.453 & 3\\
% & 0.00774 & 13 \\
% \midrule
% \multirow{2}{*}{3-regular 50Q}  & Timeout & -\\
% & 0.0292 & 34.4 \\
% \midrule
% \multirow{2}{*}{3-regular 100Q}  & Timeout & -\\
% & 0.0785 & 81.2 \\

% \midrule
% \multirow{2}{*}{4-regular 6Q}  & 18.064 & 5\\
% % \midrule
% &  0.00145 & 6.9 \\
% \midrule
% \multirow{2}{*}{4-regular 10Q}  & 3934.0 & 5\\
% & 0.00309 & 8.6 \\
% \midrule
% \multirow{2}{*}{4-regular 20Q}  & Timeout & -\\
% & 0.00873 & 17.2 \\
% \midrule
% \multirow{2}{*}{4-regular 50Q}  & Timeout & -\\
% & 0.0313 & 42.4 \\
% \midrule
% \multirow{2}{*}{4-regular 100Q}  & Timeout & -\\
% & 0.0852 & 98.1 \\

 & Benchmark & &6Q & 10Q& 20Q&50Q&100Q\\
\midrule
 \multirow{4}{*}{3-reg.} & \multirow{2}{*}{runtime(s)} & solver~\cite{tan2022} & 0.173& 0.381&74.5&{\color{red}timeout}&{\color{red}timeout}\\
& &iter-p~\cite{tan2023compiling}  &0.509 &2.16 &14.6 &966  &{\color{red}timeout} \\
& &\textbf{Ours}  &5.57E-3&9.89E-3&1.07E-2&7.52E-2&1.77E-1\\
& \multirow{2}{*}{depth} & solver~\cite{tan2022}&3&3&3&-&-\\
& &iter-p~\cite{tan2023compiling}  &3 &5 &6 &10  &- \\
& &\textbf{Ours} &5&7&11&24&45\\
\midrule
\multirow{4}{*}{4-reg.} & \multirow{2}{*}{runtime(s)} & solver~\cite{tan2022} &18.1&3.93E3&{\color{red}timeout}&{\color{red}timeout}&{\color{red}timeout}\\
& &iter-p~\cite{tan2023compiling}  &0.852 &2.64 &23.4 &2.34E3 &{\color{red}timeout} \\
&&\textbf{Ours} &6.25E-3&9.31E-3&2.10E-2&7.23E-2& 3.42E-1\\
&\multirow{2}{*}{depth}&solver~\cite{tan2022}&5&5&-&-&-\\
& &iter-p~\cite{tan2023compiling}  &5 &6 &8 &15 &- \\
&&\textbf{Ours}&6&9&15&32&60\\

% \midrule
% & 0.00131 & 6 \\

% 3-regular & 6 & 0.3811 \\
% \midrule
% 3-regular & 6 & 74.453 \\
\bottomrule
\end{tabular}

% 0.00276
% 0.00774
% 0.0292
% 0.0785

\label{tab:comp_solver}
\end{table}

\begin{figure}[t]
    \centering
    \includegraphics[width=0.96\columnwidth]{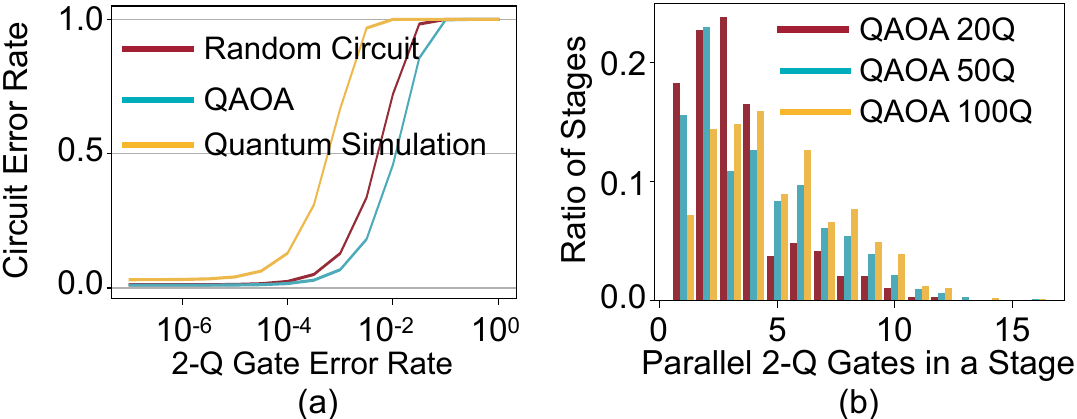}
    \caption{(a) Overall error rate vs. 2-Q gate error rate for random 6Q circuits with two 2Q gates per qubit, QAOA circuits based on random 3-regular graphs, and 5Q quantum simulation circuits with 100 Pauli strings and $p=0.1$. (b) Ratio of total stages vs number of parallel 2-Q gates in a stage using \name on QAOA circuits with 20-Q, 50-Q, and 100-Q.}
    \label{fig:2q_error_rate1}
\end{figure}

\noindent\textbf{\textit{How does the 2-Q gate error rate affect the overall error rate?}}
Fig.~\ref{fig:2q_error_rate1} (a) shows the relation between the overall error rate and the 2-Q gate error rate. We model the circuit error with the equation introduced~\cite{tan2022}, where $\epsilon$ is the overall error rate:
\begin{equation}\label{eqn:error_rate}
    \epsilon=1-f_2^{NT}f_1^{G_1}\exp\left(-N\frac{\sum_i T_0\sqrt{D_i}}{T_2}\right),
\end{equation}
$N$ is the maximum number of qubits used (including AOD and SLM), and $T$ is the circuit depth. $G_1$ is the number of 1-Q gates. $f_1$ and $f_2$ are the fidelity of 1-Q and 2-Q gates, respectively. $T_2$ is the coherence time of the qubit, and $T_0$ is the characteristic time of atom movement. $D_i$ is the maximum distance atoms moved in stage $i$. In our estimation, we choose $f_1=99.9\%$, $T_2=1.5$s, and $T_0=300\mu $s~\cite{tan2022}. The three benchmarks used here are 1) quantum simulation circuits with 5 qubits and 100 non-trivial Pauli strings with $p=0.1$, 2) random 5Q circuits with an average of two 2-Q gates per qubit, and 3) QAOA circuits for random 3-regular graphs. The error rates are below $0.5$ when the 2-Q gate has an error rate below $10^{-3}$.

\noindent\textbf{\textit{What is the distribution of the parallelism?}}
Fig.~\ref{fig:2q_error_rate1} (b) shows the percentage of stages with the number of 2-Q gates simultaneously executed for QAOA problems. The average parallelism of 20Q, 50Q, and 100Q problems are 3.32, 4.13, and 4.90, respectively. As the problem scales up, the parallelism of the problem is also increased. 

\begin{figure}[t]
    \centering
    \includegraphics[width=0.95\columnwidth]{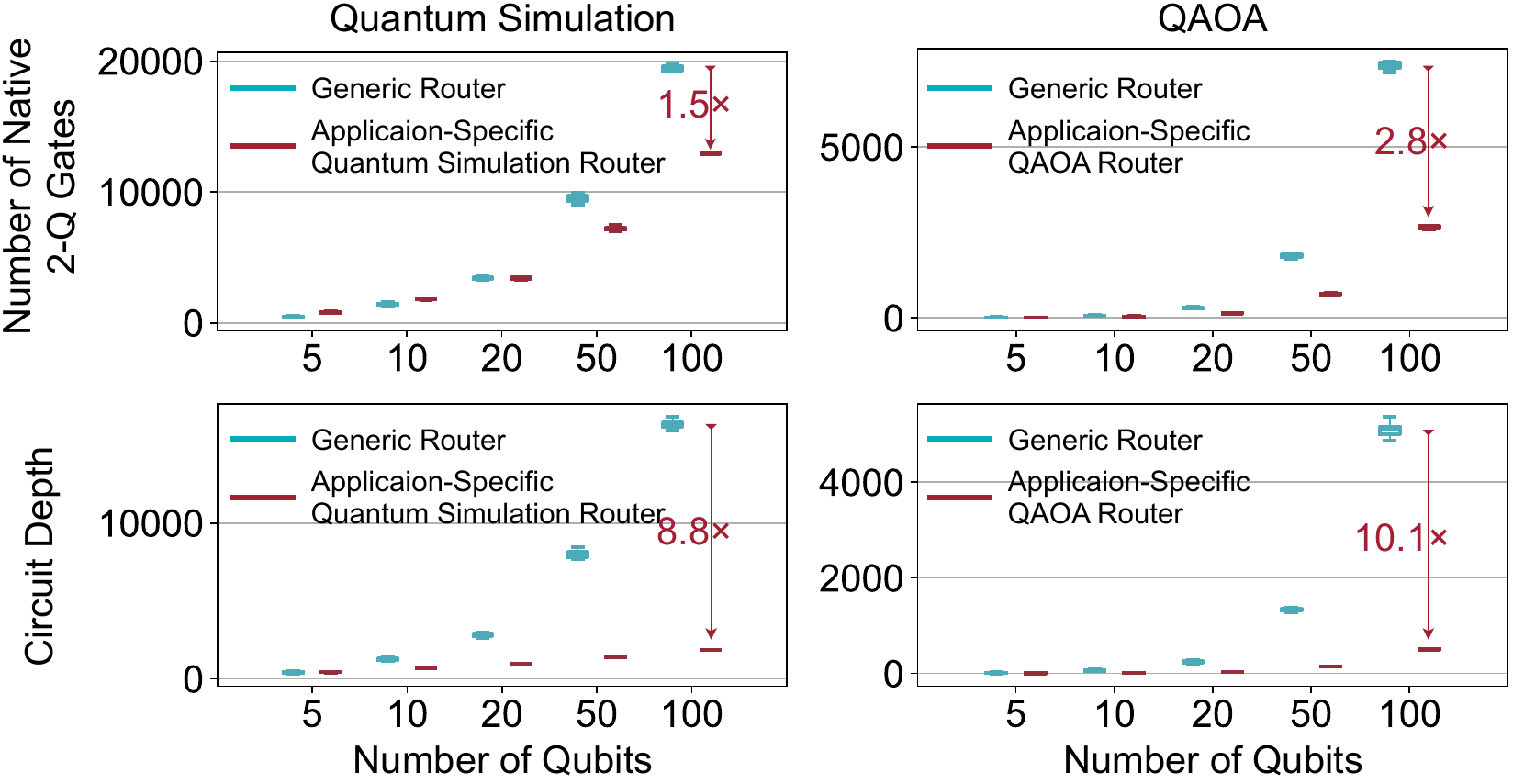}
    \caption{Advantage of our application specific Quantum Simulation and QAOA router comparing to the generic router.}
    \label{fig:specific_router}
\end{figure}

\noindent\textbf{\textit{Whether the application-specific compilers bring better performance for quantum simulation and QAOA?}} Fig.~\ref{fig:specific_router} shows the advantage of the domain-specific compiler compared to the general compiler. For quantum simulation, the domain-specific compiler reduces the 2-Q gate count by 1.5$\times$ and the circuit depth by 8.8$\times$. For QAOA, the domain-specific compiler reduces the 2-Q gate count by 2.8$\times$ and the circuit depth by 10.1$\times$. The advantages come from domain-specific heuristics that minimize the circuit depth.

\noindent\textbf{\textit{How scalable is the \name?}}
We test \name with a large number of qubits to show its scalability. For the QAOA problem, we choose random graphs with edge $p=0.5$. It takes 1.51s, 10.75s, and 129.50s to compile 500, 1000, and 2000 qubits. 
For quantum simulation problems, we choose 100 random Pauli strings. It takes 6.91s, 14.28s, and 30.48s to compile 500, 1000, and 2000 qubits. 
We generate random circuits with a depth of 10 for general circuits, and it takes 2.64s, 8.70s, and 32.31s to compile 500, 1000, and 2000 qubits. 
The fast speed proves that \name is scalable and can handle large-scale problems.

\section{Related Works}

\textbf{Compilers for Neutral Atom Arrays.} Previously, research focused on fixed atom arrays with static coupling.
Ref.~\cite{baker2021exploiting} introduced the first compiler framework for this architecture, extending existing techniques for superconducting devices and addressing unique constraints, including long-range interaction restriction zones and sporadic atom loss.
Ref.~\cite{ustc-neutral-atom} further considers gate durations in compilation.
Geyser (Ref.~\cite{geyser}) leverages native 3-Q operations by blocking 3-Q sub-circuits and re-synthesizing them. A concurrent work FPQA-C~\cite{wang2023fpqac} proposed a MAX k-Cut-based algorithm to perform qubit mapping and developed a router to schedule the gates. It does not leverage any ancilla qubits and thus is sub-optimal on circuit depth.

Ref.~\cite{iccad21-brandhoher-buchler-polian-optimal-mapping-atoms} first considered atom movement, exploring a hypothetical architecture with `1D displacement' reconfigurability, more restrictive than FPQA.
Ref.~\cite{tan2022} presented the initial FPQA compiler, formulating constraints and using an SMT solver for qubit mapping and routing.
However, the solver-based method's scalability is limited by the exponential SMT solving.
Their updated work~\cite{tan2023compiling}, sacrificing some optimality for scalability, still struggles to handle a 100-qubit circuit within a day.
Additionally, it is worth noting that they employ \textit{atom transfer} operations, moving an AOD atom to an empty SLM trap when in proximity and vice versa.
While atom transfer is already utilized in experiments \cite{ebadi_na}, frequent transfers `heat up' the atoms, potentially resulting in atom loss errors.

\textbf{Compilers for Emerging Quantum Architectures.}
Superconducting quantum computers have been widely used in quantum computing research. However, there has been a growing interest in exploring alternative quantum computing hardware, such as neutral atom machines and trapped ion machines. Our work delves into the compiler design for neutral atom systems, building upon insights from previous works~\cite{baker2021exploiting, geyser, tan2022}. Furthermore, we have observed a surge in compiler designs tailored for trapped ion systems in recent years. Notable among these are TILT~\cite{wu2020tilt}, designed for linear chain trapped ion systems; compilers for Quantum Charge Coupled Device-based trapped ion architectures as discussed in Ref.\cite{murali2020}; and compilers for shuttling-based trapped ion architectures as explored in Ref.\cite{Kreppel:2022oyr}. Additionally, a variety of compilation techniques applicable to diverse quantum computing settings have been proposed~\cite{ravi2022vaqem, 10.1145/3503222.3507703, das2021jigsaw, 10.1145/3445814.3446743, murali2019noise, tannu2019not, li2019tackling, molavi2022qubit, liuqucloud, zhang2021time, merrill2014progress, low2014optimal, brown2004arbitrarily, xie2022suppressing, hahn1950spin, viola1999dynamical, biercuk2009optimized, lidar2014review, das2021adapt, wallman2016noise, zhang2021hidden, murali2020software, wu2020tilt, versluis2017scalable, helmer2009cavity, ding2020systematic, li2022paulihedral, lao20222qan, cheng2022topgen}.
In this work, we target an emerging FPQA (Field Programmable Quantum Array) device implemented with dynamically reconfigurable atom arrays. Given the dynamic nature of the coupling map in such devices, traditional compiler techniques cannot be directly applied. To address this challenge, we introduce a novel flying ancilla-based compilation framework. This framework is designed to generate low-depth compiled circuits that maintain high scalability, catering specifically to the unique architecture of FPQA devices.

\textbf{Qubit Mapping and Instruction Scheduling.}
Since noise forms the bottleneck of NISQ machines, many noise-adaptive quantum compilation techniques have been proposed~\cite{ravi2022vaqem, 10.1145/3503222.3507703, das2021jigsaw, 10.1145/3445814.3446743, wang2022quantumnas, wang2022quantumnat, wang2022qoc, wang2022quest, wang2023robuststate, wang2023transformerqec, wang2023dgr, zheng2022sncqa}. Examples include various gate errors which can be suppressed by qubit mapping~\cite{murali2019noise, tannu2019not, li2019tackling, molavi2022qubit, liuqucloud, zhang2021time}, composite pulses~\cite{merrill2014progress, low2014optimal, brown2004arbitrarily, xie2022suppressing}, dynamical decoupling ~\cite{hahn1950spin, viola1999dynamical, biercuk2009optimized, lidar2014review, das2021adapt}, randomized compiling~\cite{wallman2016noise}, hidden inverses~\cite{zhang2021hidden}, instruction scheduling~\cite{murali2020software, wu2020tilt,zhang2023disq}, frequency tuning~\cite{versluis2017scalable, helmer2009cavity, ding2020systematic}, parallel execution on multiple machines~\cite{stein2022eqc}, algorithm-aware compilation~\cite{li2022paulihedral, lao20222qan, cheng2022topgen}, qubit specific basis gate~\cite{lin2022let, li2021software} and various pulse level optimizations~\cite{liang2022variational, liang2023pan, liang2023advantage, liang2023hybrid}. 
Qubit mapping and routing, also named quantum layout synthesis or qubit allocation/placement, has been a popular research topic in the QC community \cite{ asplos19-li-ding-xie-sabre-mapping, dac19-wille-burgholzer-zulehner-mapping-minimal-swaph, isca19-murali-linke-martonisi-abhari-nguyen-alderete-triq-architecture-studies, asplos21-zhang-hayes-qiu-jin-chen-zhang-time-optimal-mapping, iccad21-tan-cong-qubit-mapping-absorption, iccad20-tan-cong-optimal-layout-synthesis, tcad08-maslov-falconer-mosca-placement, cgo18-siraichi-santos-collange-pereira-qubit-allocation, date18-zulehner-paler-wille-efficient-mapping-ibmqx, zhou_monte_2020, DBLP:conf/micro/MolaviXDPTA22, fan-reinforcement-learning}. Instruction scheduling has also been widely explored~\cite{murali2020software, wu2020tilt, smith2021, stein2022eqc,lin2022let, li2021software}. The most relevant previous work is Brandhofer \etal~\cite{iccad21-brandhoher-buchler-polian-optimal-mapping-atoms}, which considers a hypothetical atom array architecture with `1D displacement' reconfigurability, a much more restrictive architecture than FPQA.
They construct a \textit{potential} coupling graph of all potential connections and leverage an exact router to insert SWAPs, which limits the scalability.

\section{Conclusion and Outlook}

We architect the emerging neutral atom arrays -- the field programmable quantum array (FPQA) with runtime-movable atoms.
On this highly flexible FPQA, we propose mapping all the qubits to the fixed atoms and then leveraging movable atoms as ancillas to route between fixed qubits and perform 2-Q gates.
To increase the parallelism of 2-Q gates, we design a high-parallelism generic router for arbitrary circuits and application-specific routing strategies for quantum simulation and QAOA. We hope this work can open up the avenue for future research on \fpqa.
Specifically, although the heuristics presented yield good results with short runtimes, a more general search framework where one can trade time for even higher solution quality would be valuable.
It is also exciting to explore other domains like circuits involved in quantum error correction protocols, which is the building block for future fault-tolerant QPUs made with FPQA.
Furthermore, given a scalable compiler, one can use the compilation results to guide hardware developments regarding potential features, e.g., multiple AODs or multiple `zones' in FPQA, some for Rydberg interaction and others only for storage.

\section*{Acknowledgement}
This work is partially supported by NSF grant 2313083, MIT-IBM Watson AI Lab, and Qualcomm Innovation Fellowship. The authors would like to thank Dolev Bluvstein, Mikhail D. Lukin, and Hengyun Zhou for valuable discussions on neutral atom arrays.

% \appendix
% \input{texts/6_appendix}

% \newpage

{\small
\balance
\bibliographystyle{ACM-Reference-Format}
\bibliography{main}

%%% -*-BibTeX-*-
%%% Do NOT edit. File created by BibTeX with style
%%% ACM-Reference-Format-Journals [18-Jan-2012].

\begin{thebibliography}{86}

%%% ====================================================================
%%% NOTE TO THE USER: you can override these defaults by providing
%%% customized versions of any of these macros before the \bibliography
%%% command.  Each of them MUST provide its own final punctuation,
%%% except for \shownote{}, \showDOI{}, and \showURL{}.  The latter two
%%% do not use final punctuation, in order to avoid confusing it with
%%% the Web address.
%%%
%%% To suppress output of a particular field, define its macro to expand
%%% to an empty string, or better, \unskip, like this:
%%%
%%% \newcommand{\showDOI}[1]{\unskip}   % LaTeX syntax
%%%
%%% \def \showDOI #1{\unskip}           % plain TeX syntax
%%%
%%% ====================================================================

\ifx \showCODEN    \undefined \def \showCODEN     #1{\unskip}     \fi
\ifx \showDOI      \undefined \def \showDOI       #1{#1}\fi
\ifx \showISBNx    \undefined \def \showISBNx     #1{\unskip}     \fi
\ifx \showISBNxiii \undefined \def \showISBNxiii  #1{\unskip}     \fi
\ifx \showISSN     \undefined \def \showISSN      #1{\unskip}     \fi
\ifx \showLCCN     \undefined \def \showLCCN      #1{\unskip}     \fi
\ifx \shownote     \undefined \def \shownote      #1{#1}          \fi
\ifx \showarticletitle \undefined \def \showarticletitle #1{#1}   \fi
\ifx \showURL      \undefined \def \showURL       {\relax}        \fi
% The following commands are used for tagged output and should be
% invisible to TeX
\providecommand\bibfield[2]{#2}
\providecommand\bibinfo[2]{#2}
\providecommand\natexlab[1]{#1}
\providecommand\showeprint[2][]{arXiv:#2}

\bibitem[ibm({[n.\,d.]})]%
        {ibm433}
 \bibinfo{year}{[n.\,d.]}\natexlab{}.
\newblock \bibinfo{howpublished}{\url{https://newsroom.ibm.com/2022-11-09-IBM-Unveils-400-Qubit-Plus-Quantum-Processor-and-Next-Generation-IBM-Quantum-System-Two}}.
\newblock


\bibitem[rig({[n.\,d.]})]%
        {rigetti}
 \bibinfo{year}{[n.\,d.]}\natexlab{}.
\newblock \bibinfo{howpublished}{\url{https://www.rigetti.com/}}.
\newblock


\bibitem[goo({[n.\,d.]})]%
        {google72}
 \bibinfo{year}{[n.\,d.]}\natexlab{}.
\newblock \bibinfo{howpublished}{\url{https://ai.googleblog.com/2018/03/a-preview-of-bristlecone-googles-new.html}}.
\newblock


\bibitem[int({[n.\,d.]})]%
        {intel49}
 \bibinfo{year}{[n.\,d.]}\natexlab{}.
\newblock \bibinfo{howpublished}{\url{https://spectrum.ieee.org/tech-talk/computing/hardware/intels-49qubit-chip-aims-for-quantum-supremacy}}.
\newblock


\bibitem[que({[n.\,d.]})]%
        {quera256}
 \bibinfo{year}{[n.\,d.]}\natexlab{}.
\newblock \bibinfo{howpublished}{\url{ https://www.quera.com/aquila}}.
\newblock


\bibitem[ion({[n.\,d.]})]%
        {ionq-multi-chain}
 \bibinfo{year}{[n.\,d.]}\natexlab{}.
\newblock \bibinfo{howpublished}{\url{https://ionq.com/posts/august-25-2021-deep-dive-reconfigurable-multicore-quantum-architecture}}.
\newblock


\bibitem[Alam et~al\mbox{.}(2020a)]%
        {alam2020efficient}
\bibfield{author}{\bibinfo{person}{M. Alam} {et~al\mbox{.}}} \bibinfo{year}{2020}\natexlab{a}.
\newblock \showarticletitle{An efficient circuit compilation flow for quantum approximate optimization algorithm}.
\newblock
\newblock
\shownote{DAC'20}.


\bibitem[Alam et~al\mbox{.}(2020b)]%
        {alam2020noise}
\bibfield{author}{\bibinfo{person}{M. Alam} {et~al\mbox{.}}} \bibinfo{year}{2020}\natexlab{b}.
\newblock \showarticletitle{Noise resilient compilation policies for quantum approximate optimization algorithm}.
\newblock
\newblock
\shownote{ICCAD'20}.


\bibitem[Baker et~al\mbox{.}({[n.\,d.]})]%
        {baker2021exploiting}
\bibfield{author}{\bibinfo{person}{J.~M. Baker} {et~al\mbox{.}}} \bibinfo{year}{[n.\,d.]}\natexlab{}.
\newblock \showarticletitle{Exploiting long-distance interactions and tolerating atom loss in neutral atom quantum architectures}.
\newblock
\newblock
\shownote{ISCA'21}.


\bibitem[Biercuk et~al\mbox{.}(2009)]%
        {biercuk2009optimized}
\bibfield{author}{\bibinfo{person}{Michael~J Biercuk}, \bibinfo{person}{Hermann Uys}, \bibinfo{person}{Aaron~P VanDevender}, \bibinfo{person}{Nobuyasu Shiga}, \bibinfo{person}{Wayne~M Itano}, {and} \bibinfo{person}{John~J Bollinger}.} \bibinfo{year}{2009}\natexlab{}.
\newblock \showarticletitle{Optimized dynamical decoupling in a model quantum memory}.
\newblock \bibinfo{journal}{\emph{Nature}} \bibinfo{volume}{458}, \bibinfo{number}{7241} (\bibinfo{year}{2009}), \bibinfo{pages}{996--1000}.
\newblock


\bibitem[Bluvstein et~al\mbox{.}(2022)]%
        {bluvstein2022quantum}
\bibfield{author}{\bibinfo{person}{D. Bluvstein} {et~al\mbox{.}}} \bibinfo{year}{2022}\natexlab{}.
\newblock \showarticletitle{A quantum processor based on coherent transport of entangled atom arrays}.
\newblock \bibinfo{journal}{\emph{Nature}} \bibinfo{volume}{604}, \bibinfo{number}{7906} (\bibinfo{year}{2022}), \bibinfo{pages}{451--456}.
\newblock


\bibitem[Brandhofer et~al\mbox{.}(2021)]%
        {iccad21-brandhoher-buchler-polian-optimal-mapping-atoms}
\bibfield{author}{\bibinfo{person}{S. Brandhofer} {et~al\mbox{.}}} \bibinfo{year}{2021}\natexlab{}.
\newblock \showarticletitle{Optimal mapping for near-term quantum architectures based on {Rydberg} atoms}.
\newblock
\newblock
\shownote{ICCAD'21}.


\bibitem[Brown et~al\mbox{.}(2004)]%
        {brown2004arbitrarily}
\bibfield{author}{\bibinfo{person}{Kenneth~R Brown}, \bibinfo{person}{Aram~W Harrow}, {and} \bibinfo{person}{Isaac~L Chuang}.} \bibinfo{year}{2004}\natexlab{}.
\newblock \showarticletitle{Arbitrarily accurate composite pulse sequences}.
\newblock \bibinfo{journal}{\emph{Physical Review A}} \bibinfo{volume}{70}, \bibinfo{number}{5} (\bibinfo{year}{2004}), \bibinfo{pages}{052318}.
\newblock


\bibitem[Cheng et~al\mbox{.}(2022)]%
        {cheng2022topgen}
\bibfield{author}{\bibinfo{person}{Jinglei Cheng}, \bibinfo{person}{Hanrui Wang}, \bibinfo{person}{Zhiding Liang}, \bibinfo{person}{Yiyu Shi}, \bibinfo{person}{Song Han}, {and} \bibinfo{person}{Xuehai Qian}.} \bibinfo{year}{2022}\natexlab{}.
\newblock \showarticletitle{TopGen: Topology-Aware Bottom-Up Generator for Variational Quantum Circuits}.
\newblock \bibinfo{journal}{\emph{arXiv preprint arXiv:2210.08190}} (\bibinfo{year}{2022}).
\newblock


\bibitem[Das et~al\mbox{.}(2021b)]%
        {das2021adapt}
\bibfield{author}{\bibinfo{person}{Poulami Das}, \bibinfo{person}{Swamit Tannu}, \bibinfo{person}{Siddharth Dangwal}, {and} \bibinfo{person}{Moinuddin Qureshi}.} \bibinfo{year}{2021}\natexlab{b}.
\newblock \showarticletitle{Adapt: Mitigating idling errors in qubits via adaptive dynamical decoupling}. In \bibinfo{booktitle}{\emph{MICRO-54: 54th Annual IEEE/ACM International Symposium on Microarchitecture}}. \bibinfo{pages}{950--962}.
\newblock


\bibitem[Das et~al\mbox{.}(2021a)]%
        {das2021jigsaw}
\bibfield{author}{\bibinfo{person}{Poulami Das}, \bibinfo{person}{Swamit Tannu}, {and} \bibinfo{person}{Moinuddin Qureshi}.} \bibinfo{year}{2021}\natexlab{a}.
\newblock \showarticletitle{Jigsaw: Boosting fidelity of nisq programs via measurement subsetting}. In \bibinfo{booktitle}{\emph{MICRO-54: 54th Annual IEEE/ACM International Symposium on Microarchitecture}}. \bibinfo{pages}{937--949}.
\newblock


\bibitem[Ding et~al\mbox{.}(2020)]%
        {ding2020systematic}
\bibfield{author}{\bibinfo{person}{Yongshan Ding}, \bibinfo{person}{Pranav Gokhale}, \bibinfo{person}{Sophia~Fuhui Lin}, \bibinfo{person}{Richard Rines}, \bibinfo{person}{Thomas Propson}, {and} \bibinfo{person}{Frederic~T Chong}.} \bibinfo{year}{2020}\natexlab{}.
\newblock \showarticletitle{Systematic Crosstalk Mitigation for Superconducting Qubits via Frequency-Aware Compilation}. In \bibinfo{booktitle}{\emph{2020 53rd Annual IEEE/ACM International Symposium on Microarchitecture (MICRO)}}. IEEE, \bibinfo{pages}{201--214}.
\newblock


\bibitem[Ebadi et~al\mbox{.}(2022)]%
        {ebadi_na}
\bibfield{author}{\bibinfo{person}{S. Ebadi} {et~al\mbox{.}}} \bibinfo{year}{2022}\natexlab{}.
\newblock \showarticletitle{Quantum optimization of maximum independent set using Rydberg atom arrays}.
\newblock \bibinfo{journal}{\emph{Science}} \bibinfo{volume}{376}, \bibinfo{number}{6598} (\bibinfo{year}{2022}), \bibinfo{pages}{1209--1215}.
\newblock


\bibitem[Evered et~al\mbox{.}(2023)]%
        {evered2023highfidelity}
\bibfield{author}{\bibinfo{person}{S.~J. Evered} {et~al\mbox{.}}} \bibinfo{year}{2023}\natexlab{}.
\newblock \showarticletitle{High-fidelity parallel entangling gates on a neutral atom quantum computer}.
\newblock \bibinfo{journal}{\emph{Nature}}  \bibinfo{volume}{622} (\bibinfo{year}{2023}), \bibinfo{pages}{268–272}.
\newblock


\bibitem[Fan et~al\mbox{.}(2022)]%
        {fan-reinforcement-learning}
\bibfield{author}{\bibinfo{person}{H. Fan} {et~al\mbox{.}}} \bibinfo{year}{2022}\natexlab{}.
\newblock \showarticletitle{Optimizing quantum circuit placement via machine learning}.
\newblock
\newblock
\shownote{DAC'22}.


\bibitem[Gokhale et~al\mbox{.}(2021)]%
        {gokhale-qce21-fanout}
\bibfield{author}{\bibinfo{person}{P. Gokhale} {et~al\mbox{.}}} \bibinfo{year}{2021}\natexlab{}.
\newblock \showarticletitle{Quantum fan-out: circuit optimizations and technology}.
\newblock
\newblock
\shownote{QCE'21}.


\bibitem[Graham et~al\mbox{.}(2022)]%
        {graham_multi-qubit_2022}
\bibfield{author}{\bibinfo{person}{T.~M. Graham} {et~al\mbox{.}}} \bibinfo{year}{2022}\natexlab{}.
\newblock \showarticletitle{Multi-qubit entanglement and algorithms on a neutral-atom quantum computer}.
\newblock \bibinfo{journal}{\emph{Nature}} \bibinfo{volume}{604}, \bibinfo{number}{7906} (\bibinfo{year}{2022}), \bibinfo{pages}{457--462}.
\newblock


\bibitem[Hahn(1950)]%
        {hahn1950spin}
\bibfield{author}{\bibinfo{person}{Erwin~L Hahn}.} \bibinfo{year}{1950}\natexlab{}.
\newblock \showarticletitle{Spin echoes}.
\newblock \bibinfo{journal}{\emph{Physical review}} \bibinfo{volume}{80}, \bibinfo{number}{4} (\bibinfo{year}{1950}), \bibinfo{pages}{580}.
\newblock


\bibitem[Helmer et~al\mbox{.}(2009)]%
        {helmer2009cavity}
\bibfield{author}{\bibinfo{person}{Ferdinand Helmer}, \bibinfo{person}{Matteo Mariantoni}, \bibinfo{person}{Austin~G Fowler}, \bibinfo{person}{Jan von Delft}, \bibinfo{person}{Enrique Solano}, {and} \bibinfo{person}{Florian Marquardt}.} \bibinfo{year}{2009}\natexlab{}.
\newblock \showarticletitle{Cavity grid for scalable quantum computation with superconducting circuits}.
\newblock \bibinfo{journal}{\emph{EPL (Europhysics Letters)}} \bibinfo{volume}{85}, \bibinfo{number}{5} (\bibinfo{year}{2009}), \bibinfo{pages}{50007}.
\newblock


\bibitem[H{\o}yer et~al\mbox{.}(2005)]%
        {fanout05}
\bibfield{author}{\bibinfo{person}{P. H{\o}yer} {et~al\mbox{.}}} \bibinfo{year}{2005}\natexlab{}.
\newblock \showarticletitle{Quantum fan-out is powerful}.
\newblock \bibinfo{journal}{\emph{Theory of Computing}} \bibinfo{volume}{1}, \bibinfo{number}{5} (\bibinfo{year}{2005}).
\newblock


\bibitem[Kreppel et~al\mbox{.}(2022)]%
        {Kreppel:2022oyr}
\bibfield{author}{\bibinfo{person}{Fabian Kreppel}, \bibinfo{person}{Christian Melzer}, \bibinfo{person}{Janis Wagner}, \bibinfo{person}{Janine Hilder}, \bibinfo{person}{Ulrich Poschinger}, \bibinfo{person}{Ferdinand Schmidt-Kaler}, {and} \bibinfo{person}{Andr\'e Brinkmann}.} \bibinfo{year}{2022}\natexlab{}.
\newblock \showarticletitle{{Quantum Circuit Compiler for a Shuttling-Based Trapped-Ion Quantum Computer}}.
\newblock  (\bibinfo{date}{7} \bibinfo{year}{2022}).
\newblock
\showeprint[arxiv]{2207.01964}~[quant-ph]


\bibitem[Lao and Browne(2022)]%
        {lao20222qan}
\bibfield{author}{\bibinfo{person}{Lingling Lao} {and} \bibinfo{person}{Dan~E Browne}.} \bibinfo{year}{2022}\natexlab{}.
\newblock \showarticletitle{2qan: A quantum compiler for 2-local qubit hamiltonian simulation algorithms}. In \bibinfo{booktitle}{\emph{Proceedings of the 49th Annual International Symposium on Computer Architecture}}. \bibinfo{pages}{351--365}.
\newblock


\bibitem[Li et~al\mbox{.}({[n.\,d.]})]%
        {li2019tackling}
\bibfield{author}{\bibinfo{person}{G. Li} {et~al\mbox{.}}} \bibinfo{year}{[n.\,d.]}\natexlab{}.
\newblock \showarticletitle{Tackling the qubit mapping problem for NISQ-era quantum devices}.
\newblock
\newblock
\shownote{ASPLOS'19}.


\bibitem[Li et~al\mbox{.}(2019)]%
        {asplos19-li-ding-xie-sabre-mapping}
\bibfield{author}{\bibinfo{person}{G. Li} {et~al\mbox{.}}} \bibinfo{year}{2019}\natexlab{}.
\newblock \showarticletitle{Tackling the qubit mapping problem for NISQ-era quantum}.
\newblock
\newblock
\shownote{ASPLOS'19}.


\bibitem[Li et~al\mbox{.}(2022)]%
        {li2022paulihedral}
\bibfield{author}{\bibinfo{person}{G. Li} {et~al\mbox{.}}} \bibinfo{year}{2022}\natexlab{}.
\newblock \showarticletitle{Paulihedral: a generalized block-wise compiler optimization framework for quantum simulation kernels}.
\newblock
\newblock
\shownote{ASPLOS'22}.


\bibitem[Li et~al\mbox{.}(2021)]%
        {li2021software}
\bibfield{author}{\bibinfo{person}{Gushu Li}, \bibinfo{person}{Yunong Shi}, {and} \bibinfo{person}{Ali Javadi-Abhari}.} \bibinfo{year}{2021}\natexlab{}.
\newblock \showarticletitle{Software-hardware co-optimization for computational chemistry on superconducting quantum processors}. In \bibinfo{booktitle}{\emph{2021 ACM/IEEE 48th Annual International Symposium on Computer Architecture (ISCA)}}. IEEE, \bibinfo{pages}{832--845}.
\newblock


\bibitem[Li et~al\mbox{.}(2023)]%
        {ustc-neutral-atom}
\bibfield{author}{\bibinfo{person}{Y. Li} {et~al\mbox{.}}} \bibinfo{year}{2023}\natexlab{}.
\newblock \showarticletitle{Timing-aware qubit mapping and gate scheduling adapted to neutral atom quantum computing}.
\newblock \bibinfo{journal}{\emph{TCAD}} \bibinfo{volume}{42}, \bibinfo{number}{11} (\bibinfo{year}{2023}).
\newblock


\bibitem[Liang et~al\mbox{.}(2023a)]%
        {liang2023pan}
\bibfield{author}{\bibinfo{person}{Zhiding Liang}, \bibinfo{person}{Jinglei Cheng}, \bibinfo{person}{Hang Ren}, \bibinfo{person}{Hanrui Wang}, \bibinfo{person}{Fei Hua}, \bibinfo{person}{Zhixin Song}, \bibinfo{person}{Yongshan Ding}, \bibinfo{person}{Fred Chong}, \bibinfo{person}{Song Han}, \bibinfo{person}{Yiyu Shi}, {and} \bibinfo{person}{Xuehai Qian}.} \bibinfo{year}{2023}\natexlab{a}.
\newblock \bibinfo{title}{PAN: Pulse Ansatz on NISQ Machines}.
\newblock
\newblock
\showeprint[arxiv]{2208.01215}~[quant-ph]


\bibitem[Liang et~al\mbox{.}(2023b)]%
        {liang2023advantage}
\bibfield{author}{\bibinfo{person}{Zhiding Liang}, \bibinfo{person}{Jinglei Cheng}, \bibinfo{person}{Zhixin Song}, \bibinfo{person}{Hang Ren}, \bibinfo{person}{Rui Yang}, \bibinfo{person}{Hanrui Wang}, \bibinfo{person}{Kecheng Liu}, \bibinfo{person}{Peter Kogge}, \bibinfo{person}{Tongyang Li}, \bibinfo{person}{Yongshan Ding}, {et~al\mbox{.}}} \bibinfo{year}{2023}\natexlab{b}.
\newblock \showarticletitle{Towards Advantages of Parameterized Quantum Pulses}.
\newblock \bibinfo{journal}{\emph{arXiv preprint arXiv:2304.09253}} (\bibinfo{year}{2023}).
\newblock


\bibitem[Liang et~al\mbox{.}(2023c)]%
        {liang2023hybrid}
\bibfield{author}{\bibinfo{person}{Zhiding Liang}, \bibinfo{person}{Zhixin Song}, \bibinfo{person}{Jinglei Cheng}, \bibinfo{person}{Zichang He}, \bibinfo{person}{Ji Liu}, \bibinfo{person}{Hanrui Wang}, \bibinfo{person}{Ruiyang Qin}, \bibinfo{person}{Yiru Wang}, \bibinfo{person}{Song Han}, \bibinfo{person}{Xuehai Qian}, {and} \bibinfo{person}{Yiyu Shi}.} \bibinfo{year}{2023}\natexlab{c}.
\newblock \showarticletitle{Hybrid Gate-Pulse Model for Variational Quantum Algorithms}. In \bibinfo{booktitle}{\emph{2023 60th ACM/IEEE Design Automation Conference (DAC)}}. \bibinfo{pages}{1--6}.
\newblock
\urldef\tempurl%
\url{https://doi.org/10.1109/DAC56929.2023.10247923}
\showDOI{\tempurl}


\bibitem[Liang et~al\mbox{.}(2022)]%
        {liang2022variational}
\bibfield{author}{\bibinfo{person}{Zhiding Liang}, \bibinfo{person}{Hanrui Wang}, \bibinfo{person}{Jinglei Cheng}, \bibinfo{person}{Yongshan Ding}, \bibinfo{person}{Hang Ren}, \bibinfo{person}{Xuehai Qian}, \bibinfo{person}{Song Han}, \bibinfo{person}{Weiwen Jiang}, {and} \bibinfo{person}{Yiyu Shi}.} \bibinfo{year}{2022}\natexlab{}.
\newblock \showarticletitle{Variational quantum pulse learning}.
\newblock \bibinfo{journal}{\emph{arXiv preprint arXiv:2203.17267}} (\bibinfo{year}{2022}).
\newblock


\bibitem[Lidar(2014)]%
        {lidar2014review}
\bibfield{author}{\bibinfo{person}{Daniel~A Lidar}.} \bibinfo{year}{2014}\natexlab{}.
\newblock \showarticletitle{Review of decoherence free subspaces, noiseless subsystems, and dynamical decoupling}.
\newblock \bibinfo{journal}{\emph{Adv. Chem. Phys}}  \bibinfo{volume}{154} (\bibinfo{year}{2014}), \bibinfo{pages}{295--354}.
\newblock


\bibitem[Lin et~al\mbox{.}(2022)]%
        {lin2022let}
\bibfield{author}{\bibinfo{person}{Sophia~Fuhui Lin}, \bibinfo{person}{Sara Sussman}, \bibinfo{person}{Casey Duckering}, \bibinfo{person}{Pranav~S Mundada}, \bibinfo{person}{Jonathan~M Baker}, \bibinfo{person}{Rohan~S Kumar}, \bibinfo{person}{Andrew~A Houck}, {and} \bibinfo{person}{Frederic~T Chong}.} \bibinfo{year}{2022}\natexlab{}.
\newblock \showarticletitle{Let Each Quantum Bit Choose Its Basis Gates}.
\newblock \bibinfo{journal}{\emph{arXiv preprint arXiv:2208.13380}} (\bibinfo{year}{2022}).
\newblock


\bibitem[Liu and Dou(2021)]%
        {liuqucloud}
\bibfield{author}{\bibinfo{person}{Lei Liu} {and} \bibinfo{person}{Xinglei Dou}.} \bibinfo{year}{2021}\natexlab{}.
\newblock \showarticletitle{QuCloud: A New Qubit Mapping Mechanism for Multi-programming Quantum Computing in Cloud Environment}. In \bibinfo{booktitle}{\emph{2021 IEEE International Symposium on High-Performance Computer Architecture (HPCA)}}. \bibinfo{pages}{167--178}.
\newblock
\urldef\tempurl%
\url{https://doi.org/10.1109/HPCA51647.2021.00024}
\showDOI{\tempurl}


\bibitem[Low et~al\mbox{.}(2014)]%
        {low2014optimal}
\bibfield{author}{\bibinfo{person}{Guang~Hao Low}, \bibinfo{person}{Theodore~J Yoder}, {and} \bibinfo{person}{Isaac~L Chuang}.} \bibinfo{year}{2014}\natexlab{}.
\newblock \showarticletitle{Optimal arbitrarily accurate composite pulse sequences}.
\newblock \bibinfo{journal}{\emph{Physical Review A}} \bibinfo{volume}{89}, \bibinfo{number}{2} (\bibinfo{year}{2014}), \bibinfo{pages}{022341}.
\newblock


\bibitem[Magnard et~al\mbox{.}(2020)]%
        {microwavelink}
\bibfield{author}{\bibinfo{person}{P. Magnard} {et~al\mbox{.}}} \bibinfo{year}{2020}\natexlab{}.
\newblock \showarticletitle{Microwave quantum link between superconducting circuits housed in spatially separated cryogenic systems}.
\newblock \bibinfo{journal}{\emph{PRL}}  \bibinfo{volume}{125} (\bibinfo{year}{2020}), \bibinfo{pages}{260502}.
\newblock


\bibitem[Maslov et~al\mbox{.}(2008)]%
        {tcad08-maslov-falconer-mosca-placement}
\bibfield{author}{\bibinfo{person}{D. Maslov} {et~al\mbox{.}}} \bibinfo{year}{2008}\natexlab{}.
\newblock \showarticletitle{Quantum Circuit Placement}.
\newblock \bibinfo{journal}{\emph{TCAD}} \bibinfo{volume}{27}, \bibinfo{number}{4} (\bibinfo{year}{2008}), \bibinfo{pages}{752--763}.
\newblock


\bibitem[Merrill and Brown(2014)]%
        {merrill2014progress}
\bibfield{author}{\bibinfo{person}{J~True Merrill} {and} \bibinfo{person}{Kenneth~R Brown}.} \bibinfo{year}{2014}\natexlab{}.
\newblock \showarticletitle{Progress in compensating pulse sequences for quantum computation}.
\newblock \bibinfo{journal}{\emph{Quantum Information and Computation for Chemistry}} (\bibinfo{year}{2014}), \bibinfo{pages}{241--294}.
\newblock


\bibitem[Molavi et~al\mbox{.}({[n.\,d.]})]%
        {DBLP:conf/micro/MolaviXDPTA22}
\bibfield{author}{\bibinfo{person}{Abtin Molavi} {et~al\mbox{.}}} \bibinfo{year}{[n.\,d.]}\natexlab{}.
\newblock \showarticletitle{Qubit mapping and routing via MaxSAT}.
\newblock
\newblock
\shownote{MICRO'22}.


\bibitem[Molavi et~al\mbox{.}(2022)]%
        {molavi2022qubit}
\bibfield{author}{\bibinfo{person}{Abtin Molavi}, \bibinfo{person}{Amanda Xu}, \bibinfo{person}{Martin Diges}, \bibinfo{person}{Lauren Pick}, \bibinfo{person}{Swamit Tannu}, {and} \bibinfo{person}{Aws Albarghouthi}.} \bibinfo{year}{2022}\natexlab{}.
\newblock \showarticletitle{Qubit Mapping and Routing via MaxSAT}.
\newblock \bibinfo{journal}{\emph{arXiv preprint arXiv:2208.13679}} (\bibinfo{year}{2022}).
\newblock


\bibitem[Murali et~al\mbox{.}(2019b)]%
        {isca19-murali-linke-martonisi-abhari-nguyen-alderete-triq-architecture-studies}
\bibfield{author}{\bibinfo{person}{P. Murali} {et~al\mbox{.}}} \bibinfo{year}{2019}\natexlab{b}.
\newblock \showarticletitle{Full-stack, real-system quantum computer studies: architectural comparisons and design insights}.
\newblock
\newblock
\shownote{ISCA'19}.


\bibitem[Murali et~al\mbox{.}(2019a)]%
        {murali2019noise}
\bibfield{author}{\bibinfo{person}{Prakash Murali}, \bibinfo{person}{Jonathan~M Baker}, \bibinfo{person}{Ali Javadi-Abhari}, \bibinfo{person}{Frederic~T Chong}, {and} \bibinfo{person}{Margaret Martonosi}.} \bibinfo{year}{2019}\natexlab{a}.
\newblock \showarticletitle{Noise-adaptive compiler mappings for noisy intermediate-scale quantum computers}. In \bibinfo{booktitle}{\emph{Proceedings of the Twenty-Fourth International Conference on Architectural Support for Programming Languages and Operating Systems}}. \bibinfo{pages}{1015--1029}.
\newblock


\bibitem[Murali et~al\mbox{.}(2020a)]%
        {murali2020}
\bibfield{author}{\bibinfo{person}{Prakash Murali}, \bibinfo{person}{Dripto~M. Debroy}, \bibinfo{person}{Kenneth~R. Brown}, {and} \bibinfo{person}{Margaret Martonosi}.} \bibinfo{year}{2020}\natexlab{a}.
\newblock \showarticletitle{Architecting Noisy Intermediate-Scale Trapped Ion Quantum Computers}. In \bibinfo{booktitle}{\emph{Proceedings of the ACM/IEEE 47th Annual International Symposium on Computer Architecture}} (Virtual Event) \emph{(\bibinfo{series}{ISCA '20})}. \bibinfo{publisher}{IEEE Press}, \bibinfo{pages}{529–542}.
\newblock
\showISBNx{9781728146614}
\urldef\tempurl%
\url{https://doi.org/10.1109/ISCA45697.2020.00051}
\showDOI{\tempurl}


\bibitem[Murali et~al\mbox{.}(2020b)]%
        {murali2020software}
\bibfield{author}{\bibinfo{person}{Prakash Murali}, \bibinfo{person}{David~C McKay}, \bibinfo{person}{Margaret Martonosi}, {and} \bibinfo{person}{Ali Javadi-Abhari}.} \bibinfo{year}{2020}\natexlab{b}.
\newblock \showarticletitle{Software mitigation of crosstalk on noisy intermediate-scale quantum computers}. In \bibinfo{booktitle}{\emph{Proceedings of the Twenty-Fifth International Conference on Architectural Support for Programming Languages and Operating Systems}}. \bibinfo{pages}{1001--1016}.
\newblock


\bibitem[Norcia et~al\mbox{.}({[n.\,d.]})]%
        {atomcomputing}
\bibfield{author}{\bibinfo{person}{M.~A. Norcia} {et~al\mbox{.}}} \bibinfo{year}{[n.\,d.]}\natexlab{}.
\newblock \bibinfo{title}{Iterative assembly of $^{171}$Yb atom arrays in cavity-enhanced optical lattices}.
\newblock
\newblock
\newblock
\shownote{arXiv:2401.16177}.


\bibitem[Park et~al\mbox{.}(2022)]%
        {park22dac}
\bibfield{author}{\bibinfo{person}{S. Park} {et~al\mbox{.}}} \bibinfo{year}{2022}\natexlab{}.
\newblock \showarticletitle{A fast and scalable qubit-mapping method for noisy intermediate-scale quantum computers}.
\newblock
\newblock
\shownote{DAC'22}.


\bibitem[Patel et~al\mbox{.}(2022)]%
        {geyser}
\bibfield{author}{\bibinfo{person}{T. Patel} {et~al\mbox{.}}} \bibinfo{year}{2022}\natexlab{}.
\newblock \showarticletitle{Geyser: a compilation framework for quantum computing with neutral atoms}.
\newblock
\newblock
\shownote{ISCA '22}.


\bibitem[Patel and Tiwari(2021)]%
        {10.1145/3445814.3446743}
\bibfield{author}{\bibinfo{person}{Tirthak Patel} {and} \bibinfo{person}{Devesh Tiwari}.} \bibinfo{year}{2021}\natexlab{}.
\newblock \showarticletitle{Qraft: Reverse Your Quantum Circuit and Know the Correct Program Output}. In \bibinfo{booktitle}{\emph{Proceedings of the 26th ACM International Conference on Architectural Support for Programming Languages and Operating Systems}} (Virtual, USA) \emph{(\bibinfo{series}{ASPLOS '21})}. \bibinfo{publisher}{Association for Computing Machinery}, \bibinfo{address}{New York, NY, USA}, \bibinfo{pages}{443–455}.
\newblock
\showISBNx{9781450383172}
\urldef\tempurl%
\url{https://doi.org/10.1145/3445814.3446743}
\showDOI{\tempurl}


\bibitem[Ravi et~al\mbox{.}(2022)]%
        {ravi2022vaqem}
\bibfield{author}{\bibinfo{person}{Gokul~Subramanian Ravi}, \bibinfo{person}{Kaitlin~N Smith}, \bibinfo{person}{Pranav Gokhale}, \bibinfo{person}{Andrea Mari}, \bibinfo{person}{Nathan Earnest}, \bibinfo{person}{Ali Javadi-Abhari}, {and} \bibinfo{person}{Frederic~T Chong}.} \bibinfo{year}{2022}\natexlab{}.
\newblock \showarticletitle{Vaqem: A variational approach to quantum error mitigation}. In \bibinfo{booktitle}{\emph{2022 IEEE International Symposium on High-Performance Computer Architecture (HPCA)}}. IEEE, \bibinfo{pages}{288--303}.
\newblock


\bibitem[Saffman(2016)]%
        {Saffman_2016}
\bibfield{author}{\bibinfo{person}{M Saffman}.} \bibinfo{year}{2016}\natexlab{}.
\newblock \showarticletitle{Quantum computing with atomic qubits and Rydberg interactions: progress and challenges}.
\newblock \bibinfo{journal}{\emph{Journal of Physics B}} \bibinfo{volume}{49}, \bibinfo{number}{20} (\bibinfo{date}{oct} \bibinfo{year}{2016}), \bibinfo{pages}{202001}.
\newblock


\bibitem[Siraichi et~al\mbox{.}(2018)]%
        {cgo18-siraichi-santos-collange-pereira-qubit-allocation}
\bibfield{author}{\bibinfo{person}{M.~Y. Siraichi} {et~al\mbox{.}}} \bibinfo{year}{2018}\natexlab{}.
\newblock \showarticletitle{Qubit allocation}.
\newblock
\newblock
\shownote{CGO'18}.


\bibitem[Smith et~al\mbox{.}(2021)]%
        {smith2021}
\bibfield{author}{\bibinfo{person}{Kaitlin~N. Smith}, \bibinfo{person}{Gokul~Subramanian Ravi}, \bibinfo{person}{Prakash Murali}, \bibinfo{person}{Jonathan~M. Baker}, \bibinfo{person}{Nathan Earnest}, \bibinfo{person}{Ali Javadi-Abhari}, {and} \bibinfo{person}{Frederic~T. Chong}.} \bibinfo{year}{2021}\natexlab{}.
\newblock \bibinfo{title}{Error Mitigation in Quantum Computers through Instruction Scheduling}.
\newblock
\newblock
\urldef\tempurl%
\url{https://doi.org/10.48550/ARXIV.2105.01760}
\showDOI{\tempurl}


\bibitem[Stein et~al\mbox{.}(2022)]%
        {stein2022eqc}
\bibfield{author}{\bibinfo{person}{Samuel Stein}, \bibinfo{person}{Nathan Wiebe}, \bibinfo{person}{Yufei Ding}, \bibinfo{person}{Peng Bo}, \bibinfo{person}{Karol Kowalski}, \bibinfo{person}{Nathan Baker}, \bibinfo{person}{James Ang}, {and} \bibinfo{person}{Ang Li}.} \bibinfo{year}{2022}\natexlab{}.
\newblock \showarticletitle{EQC: ensembled quantum computing for variational quantum algorithms}. In \bibinfo{booktitle}{\emph{Proceedings of the 49th Annual International Symposium on Computer Architecture}}. \bibinfo{pages}{59--71}.
\newblock


\bibitem[Tan et~al\mbox{.}(2020)]%
        {iccad20-tan-cong-optimal-layout-synthesis}
\bibfield{author}{\bibinfo{person}{B. Tan} {et~al\mbox{.}}} \bibinfo{year}{2020}\natexlab{}.
\newblock \showarticletitle{Optimal layout synthesis for quantum computing}.
\newblock
\newblock
\shownote{ICCAD'20}.


\bibitem[Tan et~al\mbox{.}(2021)]%
        {iccad21-tan-cong-qubit-mapping-absorption}
\bibfield{author}{\bibinfo{person}{B. Tan} {et~al\mbox{.}}} \bibinfo{year}{2021}\natexlab{}.
\newblock \showarticletitle{Optimal qubit mapping with simultaneous gate absorption}.
\newblock
\newblock
\shownote{ICCAD'21}.


\bibitem[Tan et~al\mbox{.}(2022)]%
        {tan2022}
\bibfield{author}{\bibinfo{person}{B. Tan} {et~al\mbox{.}}} \bibinfo{year}{2022}\natexlab{}.
\newblock \showarticletitle{Qubit mapping for reconfigurable atom arrays}.
\newblock  (\bibinfo{year}{2022}).
\newblock
\newblock
\shownote{ICCAD'22}.


\bibitem[Tan et~al\mbox{.}(2024)]%
        {tan2023compiling}
\bibfield{author}{\bibinfo{person}{D.~B. Tan} {et~al\mbox{.}}} \bibinfo{year}{2024}\natexlab{}.
\newblock \showarticletitle{Compiling quantum circuits for dynamically field-programmable neutral atoms array processors}.
\newblock \bibinfo{journal}{\emph{Quantum}}  \bibinfo{volume}{8} (\bibinfo{year}{2024}).
\newblock


\bibitem[Tannu et~al\mbox{.}(2022)]%
        {10.1145/3503222.3507703}
\bibfield{author}{\bibinfo{person}{Swamit Tannu}, \bibinfo{person}{Poulami Das}, \bibinfo{person}{Ramin Ayanzadeh}, {and} \bibinfo{person}{Moinuddin Qureshi}.} \bibinfo{year}{2022}\natexlab{}.
\newblock \showarticletitle{HAMMER: Boosting Fidelity of Noisy Quantum Circuits by Exploiting Hamming Behavior of Erroneous Outcomes}. In \bibinfo{booktitle}{\emph{Proceedings of the 27th ACM International Conference on Architectural Support for Programming Languages and Operating Systems}} (Lausanne, Switzerland) \emph{(\bibinfo{series}{ASPLOS '22})}. \bibinfo{publisher}{Association for Computing Machinery}, \bibinfo{address}{New York, NY, USA}, \bibinfo{pages}{529–540}.
\newblock
\showISBNx{9781450392051}
\urldef\tempurl%
\url{https://doi.org/10.1145/3503222.3507703}
\showDOI{\tempurl}


\bibitem[Tannu and Qureshi(2019)]%
        {tannu2019not}
\bibfield{author}{\bibinfo{person}{Swamit~S Tannu} {and} \bibinfo{person}{Moinuddin~K Qureshi}.} \bibinfo{year}{2019}\natexlab{}.
\newblock \showarticletitle{Not all qubits are created equal: a case for variability-aware policies for NISQ-era quantum computers}. In \bibinfo{booktitle}{\emph{Proceedings of the Twenty-Fourth International Conference on Architectural Support for Programming Languages and Operating Systems}}. \bibinfo{pages}{987--999}.
\newblock


\bibitem[Versluis et~al\mbox{.}(2017)]%
        {versluis2017scalable}
\bibfield{author}{\bibinfo{person}{Richard Versluis}, \bibinfo{person}{Stefano Poletto}, \bibinfo{person}{Nader Khammassi}, \bibinfo{person}{Brian Tarasinski}, \bibinfo{person}{Nadia Haider}, \bibinfo{person}{David~J Michalak}, \bibinfo{person}{Alessandro Bruno}, \bibinfo{person}{Koen Bertels}, {and} \bibinfo{person}{Leonardo DiCarlo}.} \bibinfo{year}{2017}\natexlab{}.
\newblock \showarticletitle{Scalable quantum circuit and control for a superconducting surface code}.
\newblock \bibinfo{journal}{\emph{Physical Review Applied}} \bibinfo{volume}{8}, \bibinfo{number}{3} (\bibinfo{year}{2017}), \bibinfo{pages}{034021}.
\newblock


\bibitem[Viola et~al\mbox{.}(1999)]%
        {viola1999dynamical}
\bibfield{author}{\bibinfo{person}{Lorenza Viola}, \bibinfo{person}{Emanuel Knill}, {and} \bibinfo{person}{Seth Lloyd}.} \bibinfo{year}{1999}\natexlab{}.
\newblock \showarticletitle{Dynamical decoupling of open quantum systems}.
\newblock \bibinfo{journal}{\emph{Physical Review Letters}} \bibinfo{volume}{82}, \bibinfo{number}{12} (\bibinfo{year}{1999}), \bibinfo{pages}{2417}.
\newblock


\bibitem[Wallman and Emerson(2016)]%
        {wallman2016noise}
\bibfield{author}{\bibinfo{person}{Joel~J Wallman} {and} \bibinfo{person}{Joseph Emerson}.} \bibinfo{year}{2016}\natexlab{}.
\newblock \showarticletitle{Noise tailoring for scalable quantum computation via randomized compiling}.
\newblock \bibinfo{journal}{\emph{Physical Review A}} \bibinfo{volume}{94}, \bibinfo{number}{5} (\bibinfo{year}{2016}), \bibinfo{pages}{052325}.
\newblock


\bibitem[Wang et~al\mbox{.}(2022a)]%
        {wang2022quantumnas}
\bibfield{author}{\bibinfo{person}{Hanrui Wang}, \bibinfo{person}{Yongshan Ding}, \bibinfo{person}{Jiaqi Gu}, \bibinfo{person}{Yujun Lin}, \bibinfo{person}{David~Z Pan}, \bibinfo{person}{Frederic~T Chong}, {and} \bibinfo{person}{Song Han}.} \bibinfo{year}{2022}\natexlab{a}.
\newblock \showarticletitle{Quantumnas: Noise-adaptive search for robust quantum circuits}. In \bibinfo{booktitle}{\emph{2022 IEEE International Symposium on High-Performance Computer Architecture (HPCA)}}. IEEE, \bibinfo{pages}{692--708}.
\newblock


\bibitem[Wang et~al\mbox{.}(2022b)]%
        {wang2022quantumnat}
\bibfield{author}{\bibinfo{person}{Hanrui Wang}, \bibinfo{person}{Jiaqi Gu}, \bibinfo{person}{Yongshan Ding}, \bibinfo{person}{Zirui Li}, \bibinfo{person}{Frederic~T Chong}, \bibinfo{person}{David~Z Pan}, {and} \bibinfo{person}{Song Han}.} \bibinfo{year}{2022}\natexlab{b}.
\newblock \showarticletitle{QuantumNAT: quantum noise-aware training with noise injection, quantization and normalization}. In \bibinfo{booktitle}{\emph{Proceedings of the 59th ACM/IEEE Design Automation Conference}}. \bibinfo{pages}{1--6}.
\newblock


\bibitem[Wang et~al\mbox{.}(2022c)]%
        {wang2022qoc}
\bibfield{author}{\bibinfo{person}{Hanrui Wang}, \bibinfo{person}{Zirui Li}, \bibinfo{person}{Jiaqi Gu}, \bibinfo{person}{Yongshan Ding}, \bibinfo{person}{David~Z Pan}, {and} \bibinfo{person}{Song Han}.} \bibinfo{year}{2022}\natexlab{c}.
\newblock \showarticletitle{QOC: quantum on-chip training with parameter shift and gradient pruning}. In \bibinfo{booktitle}{\emph{Proceedings of the 59th ACM/IEEE Design Automation Conference}}. \bibinfo{pages}{655--660}.
\newblock


\bibitem[Wang et~al\mbox{.}(2022d)]%
        {wang2022quest}
\bibfield{author}{\bibinfo{person}{Hanrui Wang}, \bibinfo{person}{Pengyu Liu}, \bibinfo{person}{Jinglei Cheng}, \bibinfo{person}{Zhiding Liang}, \bibinfo{person}{Jiaqi Gu}, \bibinfo{person}{Zirui Li}, \bibinfo{person}{Yongshan Ding}, \bibinfo{person}{Weiwen Jiang}, \bibinfo{person}{Yiyu Shi}, \bibinfo{person}{Xuehai Qian}, {et~al\mbox{.}}} \bibinfo{year}{2022}\natexlab{d}.
\newblock \showarticletitle{Quest: Graph transformer for quantum circuit reliability estimation}. In \bibinfo{booktitle}{\emph{IEEE/ACM International Conference on Computer-Aided Design (ICCAD)}}.
\newblock


\bibitem[Wang et~al\mbox{.}(2023a)]%
        {wang2023dgr}
\bibfield{author}{\bibinfo{person}{Hanrui Wang}, \bibinfo{person}{Pengyu Liu}, \bibinfo{person}{Yilian Liu}, \bibinfo{person}{Jiaqi Gu}, \bibinfo{person}{Jonathan Baker}, \bibinfo{person}{Frederic~T. Chong}, {and} \bibinfo{person}{Song Han}.} \bibinfo{year}{2023}\natexlab{a}.
\newblock \showarticletitle{{DGR: Tackling Drifted and Correlated Noise in Quantum Error Correction via Decoding Graph Re-weighting}}. In \bibinfo{booktitle}{\emph{arXiv preprint}}.
\newblock


\bibitem[Wang et~al\mbox{.}(2023c)]%
        {wang2023transformerqec}
\bibfield{author}{\bibinfo{person}{Hanrui Wang}, \bibinfo{person}{Pengyu Liu}, \bibinfo{person}{Kevin Shao}, \bibinfo{person}{Dantong Li}, \bibinfo{person}{Jiaqi Gu}, \bibinfo{person}{David~Z Pan}, \bibinfo{person}{Yongshan Ding}, {and} \bibinfo{person}{Song Han}.} \bibinfo{year}{2023}\natexlab{c}.
\newblock \showarticletitle{TransformerQEC: Transferable Transformer for Quantum Error Correction Code Decoding}. In \bibinfo{booktitle}{\emph{IEEE/ACM International Conference on Computer-Aided Design (ICCAD), FastML for Science Workshop}}.
\newblock


\bibitem[Wang et~al\mbox{.}(2023d)]%
        {wang2023fpqac}
\bibfield{author}{\bibinfo{person}{Hanrui Wang}, \bibinfo{person}{Pengyu Liu}, \bibinfo{person}{Bochen Tan}, \bibinfo{person}{Yilian Liu}, \bibinfo{person}{Jiaqi Gu}, \bibinfo{person}{David~Z. Pan}, \bibinfo{person}{Jason Cong}, \bibinfo{person}{Umut Acar}, {and} \bibinfo{person}{Song Han}.} \bibinfo{year}{2023}\natexlab{d}.
\newblock \showarticletitle{{FPQA-C : A Compilation Framework for Field Programmable Qubit Array}}. In \bibinfo{booktitle}{\emph{arXiv preprint}}.
\newblock


\bibitem[Wang et~al\mbox{.}(2023b)]%
        {wang2023robuststate}
\bibfield{author}{\bibinfo{person}{Hanrui Wang}, \bibinfo{person}{Yilian Liu}, \bibinfo{person}{Pengyu Liu}, \bibinfo{person}{Jiaqi Gu}, \bibinfo{person}{Zirui Li}, \bibinfo{person}{Zhiding Liang}, \bibinfo{person}{Jinglei Cheng}, \bibinfo{person}{Yongshan Ding}, \bibinfo{person}{Xuehai Qian}, \bibinfo{person}{Yiyu Shi}, \bibinfo{person}{David~Z. Pan}, \bibinfo{person}{Frederic~T. Chong}, {and} \bibinfo{person}{Song Han}.} \bibinfo{year}{2023}\natexlab{b}.
\newblock \showarticletitle{RobustState: Boosting Fidelity of Quantum State Preparation via Noise-Aware Variational Training}.
\newblock \bibinfo{journal}{\emph{arXiv preprint}} (\bibinfo{year}{2023}).
\newblock


\bibitem[Wille et~al\mbox{.}(2019)]%
        {dac19-wille-burgholzer-zulehner-mapping-minimal-swaph}
\bibfield{author}{\bibinfo{person}{R. Wille} {et~al\mbox{.}}} \bibinfo{year}{2019}\natexlab{}.
\newblock \showarticletitle{Mapping quantum circuits to {IBM QX} architectures using the minimal number of {SWAP} and {H} operations}.
\newblock
\newblock
\shownote{DAC'19}.


\bibitem[Wu et~al\mbox{.}(2022)]%
        {wu22iccad}
\bibfield{author}{\bibinfo{person}{T.-A. Wu} {et~al\mbox{.}}} \bibinfo{year}{2022}\natexlab{}.
\newblock \showarticletitle{A robust quantum layout synthesis algorithm with a qubit mapping checker.}
\newblock
\newblock
\shownote{ICCAD'22}.


\bibitem[Wu et~al\mbox{.}(2020)]%
        {wu2020tilt}
\bibfield{author}{\bibinfo{person}{Xin-Chuan Wu}, \bibinfo{person}{Dripto~M Debroy}, \bibinfo{person}{Yongshan Ding}, \bibinfo{person}{Jonathan~M Baker}, \bibinfo{person}{Yuri Alexeev}, \bibinfo{person}{Kenneth~R Brown}, {and} \bibinfo{person}{Frederic~T Chong}.} \bibinfo{year}{2020}\natexlab{}.
\newblock \showarticletitle{TILT: Achieving Higher Fidelity on a Trapped-Ion Linear-Tape Quantum Computing Architecture}.
\newblock \bibinfo{journal}{\emph{arXiv preprint arXiv:2010.15876}} (\bibinfo{year}{2020}).
\newblock


\bibitem[Xie et~al\mbox{.}(2022)]%
        {xie2022suppressing}
\bibfield{author}{\bibinfo{person}{Lei Xie}, \bibinfo{person}{Jidong Zhai}, \bibinfo{person}{ZhenXing Zhang}, \bibinfo{person}{Jonathan Allcock}, \bibinfo{person}{Shengyu Zhang}, {and} \bibinfo{person}{Yi-Cong Zheng}.} \bibinfo{year}{2022}\natexlab{}.
\newblock \showarticletitle{Suppressing ZZ crosstalk of Quantum computers through pulse and scheduling co-optimization}. In \bibinfo{booktitle}{\emph{Proceedings of the 27th ACM International Conference on Architectural Support for Programming Languages and Operating Systems}}. \bibinfo{pages}{499--513}.
\newblock


\bibitem[Zhang et~al\mbox{.}(2021c)]%
        {zhang2021hidden}
\bibfield{author}{\bibinfo{person}{Bichen Zhang}, \bibinfo{person}{Swarnadeep Majumder}, \bibinfo{person}{Pak~Hong Leung}, \bibinfo{person}{Stephen Crain}, \bibinfo{person}{Ye Wang}, \bibinfo{person}{Chao Fang}, \bibinfo{person}{Dripto~M Debroy}, \bibinfo{person}{Jungsang Kim}, {and} \bibinfo{person}{Kenneth~R Brown}.} \bibinfo{year}{2021}\natexlab{c}.
\newblock \showarticletitle{Hidden Inverses: Coherent Error Cancellation at the Circuit Level}.
\newblock \bibinfo{journal}{\emph{arXiv preprint arXiv:2104.01119}} (\bibinfo{year}{2021}).
\newblock


\bibitem[Zhang et~al\mbox{.}(2021a)]%
        {zhang2021time}
\bibfield{author}{\bibinfo{person}{C. Zhang} {et~al\mbox{.}}} \bibinfo{year}{2021}\natexlab{a}.
\newblock \showarticletitle{Time-optimal qubit mapping}.
\newblock
\newblock
\shownote{ASPLOS'21}.


\bibitem[Zhang et~al\mbox{.}(2021b)]%
        {asplos21-zhang-hayes-qiu-jin-chen-zhang-time-optimal-mapping}
\bibfield{author}{\bibinfo{person}{C. Zhang} {et~al\mbox{.}}} \bibinfo{year}{2021}\natexlab{b}.
\newblock \showarticletitle{Time-optimal qubit mapping}.
\newblock
\newblock
\shownote{ASPOLS'21}.


\bibitem[Zhang et~al\mbox{.}(2023)]%
        {zhang2023disq}
\bibfield{author}{\bibinfo{person}{Junyao Zhang}, \bibinfo{person}{Hanrui Wang}, \bibinfo{person}{Gokul~Subramanian Ravi}, \bibinfo{person}{Frederic~T Chong}, \bibinfo{person}{Song Han}, \bibinfo{person}{Frank Mueller}, {and} \bibinfo{person}{Yiran Chen}.} \bibinfo{year}{2023}\natexlab{}.
\newblock \showarticletitle{DISQ: Dynamic Iteration Skipping for Variational Quantum Algorithms}. In \bibinfo{booktitle}{\emph{2023 IEEE International Conference on Quantum Computing and Engineering (QCE)}}.
\newblock


\bibitem[Zheng et~al\mbox{.}(2023)]%
        {zheng2022sncqa}
\bibfield{author}{\bibinfo{person}{Han Zheng}, \bibinfo{person}{Gokul~Subramanian Ravi}, \bibinfo{person}{Hanrui Wang}, \bibinfo{person}{Kanav Setia}, \bibinfo{person}{Frederic~T Chong}, {and} \bibinfo{person}{Junyu Liu}.} \bibinfo{year}{2023}\natexlab{}.
\newblock \showarticletitle{SnCQA: A hardware-efficient equivariant quantum convolutional circuit architecture}. In \bibinfo{booktitle}{\emph{2023 IEEE International Conference on Quantum Computing and Engineering (QCE)}}.
\newblock


\bibitem[Zhou et~al\mbox{.}(2020)]%
        {zhou_monte_2020}
\bibfield{author}{\bibinfo{person}{X. Zhou} {et~al\mbox{.}}} \bibinfo{year}{2020}\natexlab{}.
\newblock \showarticletitle{A Monte Carlo tree search framework for quantum circuit transformation}.
\newblock
\newblock
\shownote{ICCAD'20}.


\bibitem[Zulehner et~al\mbox{.}(2018)]%
        {date18-zulehner-paler-wille-efficient-mapping-ibmqx}
\bibfield{author}{\bibinfo{person}{A. Zulehner} {et~al\mbox{.}}} \bibinfo{year}{2018}\natexlab{}.
\newblock \showarticletitle{Efficient mapping of quantum circuits to the {IBM QX} architectures}.
\newblock
\newblock
\shownote{DATE'18}.


\end{thebibliography}
}

\end{document}